\documentclass[aps,pra,twocolumn,superscriptaddress,amsmath,amssymb,tightenlines,footinbib]{revtex4-1}

\usepackage{times}
\usepackage[dvipdfmx]{graphicx} 
\usepackage{bmpsize}
\usepackage[latin1]{inputenc}
\usepackage{mathrsfs}
\usepackage[dvipsnames,usenames]{color}
\usepackage{soul}
\usepackage[colorlinks,bookmarks=true,citecolor=blue,linkcolor=blue,urlcolor=blue, breaklinks=true]{hyperref}
\graphicspath{{./figs/}}

\usepackage[percent]{overpic}

\usepackage{amssymb}
\usepackage{latexsym}
\usepackage{amsmath}

\usepackage{xcolor}
\usepackage{graphicx}
\usepackage{amsfonts}
\usepackage{tabularx}
\usepackage{stmaryrd}
\usepackage{bbold}
\usepackage{array}
\usepackage{mathtools}
\usepackage{subcaption}
\usepackage{comment}

\usepackage{ragged2e} 
\DeclareCaptionJustification{justified}{\justifying}

\begin{document}

\title{Adaptive-weighted Tree Tensor Networks for disordered quantum many-body systems}
\author{Giovanni Ferrari} 
\affiliation{Dipartimento di Fisica e Astronomia ``G. Galilei'', Universit\`a di Padova, I-35131 Padova, Italy.}
\affiliation{Institut f\"ur Theoretische Physik und IQST, Universit\"at Ulm, Albert-Einstein-Allee 11, D-89069 Ulm, Germany.}
\author{Giuseppe Magnifico} 
\affiliation{Dipartimento di Fisica e Astronomia ``G. Galilei'', Universit\`a di Padova, I-35131 Padova, Italy.}
\affiliation{Padua Quantum Technologies Research Center, Universit\`a degli Studi di Padova.}
\affiliation{Istituto Nazionale di Fisica Nucleare (INFN), Sezione di Padova, I-35131 Padova, Italy.}   
\author{Simone Montangero}
\affiliation{Dipartimento di Fisica e Astronomia ``G. Galilei'', Universit\`a di Padova, I-35131 Padova, Italy.}
\affiliation{Padua Quantum Technologies Research Center, Universit\`a degli Studi di Padova.}
\affiliation{Istituto Nazionale di Fisica Nucleare (INFN), Sezione di Padova, I-35131 Padova, Italy.}
\date{\today}
\begin{abstract}
We introduce an adaptive-weighted Tree Tensor Network, for the study of disordered and inhomogeneous quantum many-body systems. This ansatz is assembled on the basis of the random couplings of the physical system with a procedure that considers a tunable weight parameter to prevent completely unbalanced trees. Using this approach, we compute the ground state of the two-dimensional quantum Ising model in the presence of quenched random disorder and frustration, with lattice size up to $32 \times 32$. We compare the results with the ones obtained using the standard homogeneous Tree Tensor Networks and the completely self-assembled Tree Tensor Networks, demonstrating a clear improvement of numerical precision as a function of the weight parameter, especially for large system sizes.
\end{abstract}
 
\maketitle

\section{Introduction} 
Quantum many-body systems and materials in real life usually present some degree of imperfection due to the presence of impurities and defects that arise, e.g., during the experimental implementations and the sample realization processes. These imperfections can be described by introducing some types of randomness or disorder in the physical modeling of the system \cite{Vojta2019}. On the one hand, disorder can degrade the quantum properties of matter by altering the distinctive features of quantum phases at equilibrium and preventing phases transitions. For instance, first-order phase transitions could be remarkably attenuated in the presence of disorder, whereas critical exponents and critical points of continuous phase transitions can be significantly modified due to the effects of random interactions \cite{Vojta_2006, Vojta_2010}. On the other hand, disorder can generate physical phenomena of great interest for both theoretical and experimental research. For instance, disorder can lead to peculiar phases in quantum spin models and magnetic systems, such as the spin glasses, in which spins are aligned randomly, without the formation of a regular pattern \cite{Rieger, Spin_Glass_1, Spin_Glass_2, Mossi_2017}. Also, disorder stands at the basis of Anderson localization, that is, the space localization of the quantum states of particles \cite{PhysRev.109.1492, doi:10.1142/7663}. This intriguing phenomenon is specifically due to disorder since it is observable even when the particles do not feel any interactions. The study of this important effect led to fundamental results in our understanding of the static and dynamical properties of conductive and insulating materials and the peculiar features of the metal-insulator quantum phase transitions \cite{RevModPhys.80.1355, RevModPhys.66.261}. The presence of disorder combined with strong interactions may also lead to the intriguing effect of many-body localization, resulting in non-thermalizing phases of matter and logarithmic growth of entanglement for isolated systems \cite{many_body_localization_1, Abanin_2019, De_Luca_2013}. Furthermore, disorder 
lies at the basis of the phenomenon of entanglement enhancement \cite{PhysRevB.74.174204, PhysRevB.102.214201, PhysRevB.103.024202} and plays a key role in the study of complex atomic and molecular systems relevant to quantum chemistry \cite{disorder_quantum_chem}.

\begin{figure}
\begin{subfigure}{.5\textwidth}
\caption{}
  \centering
  \includegraphics[width=.9\linewidth]{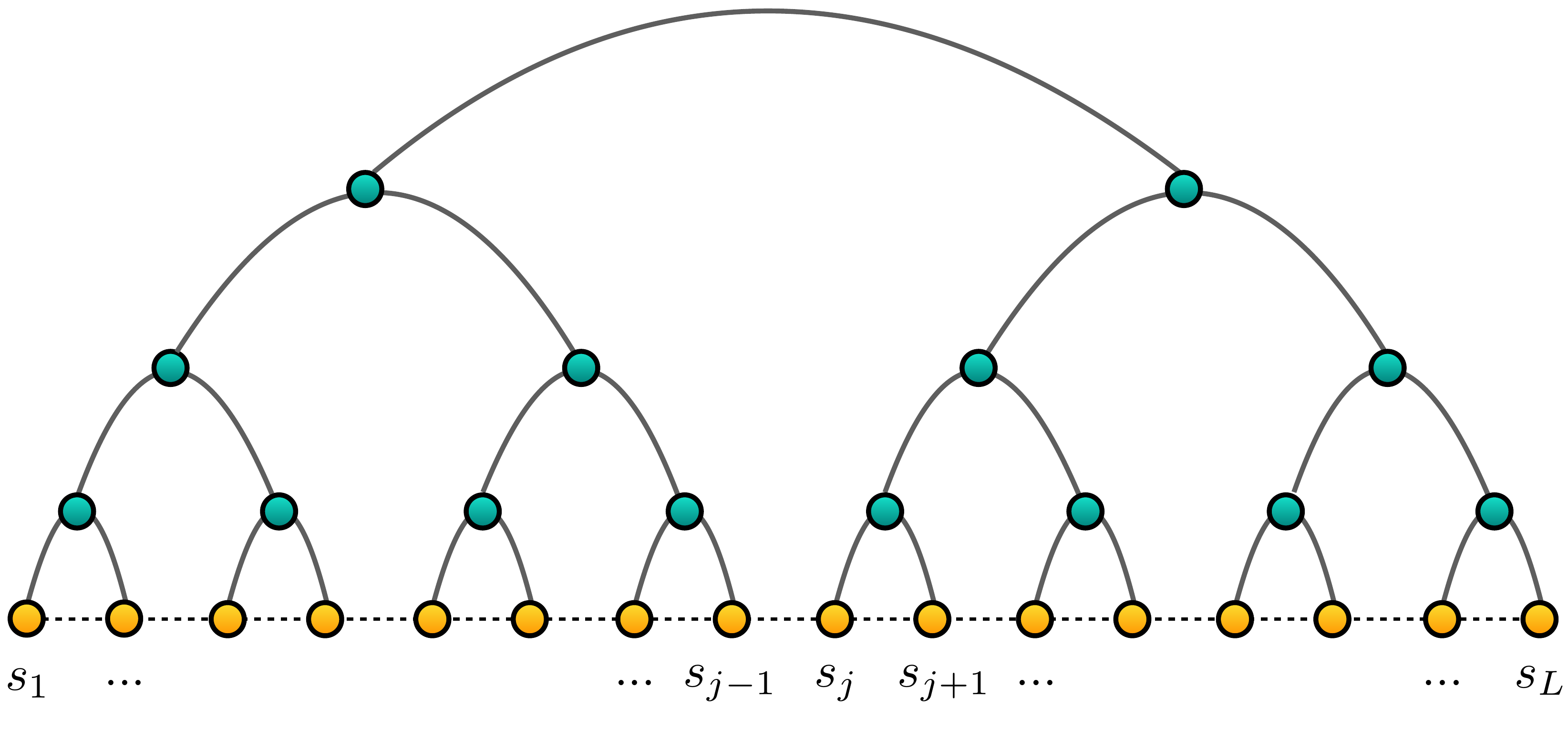}  
  \label{fig:binary_tree}
  \newline
\end{subfigure}
\newline
\begin{subfigure}{.5\textwidth}
  \caption{}
  \centering
  \includegraphics[width=.9\linewidth]{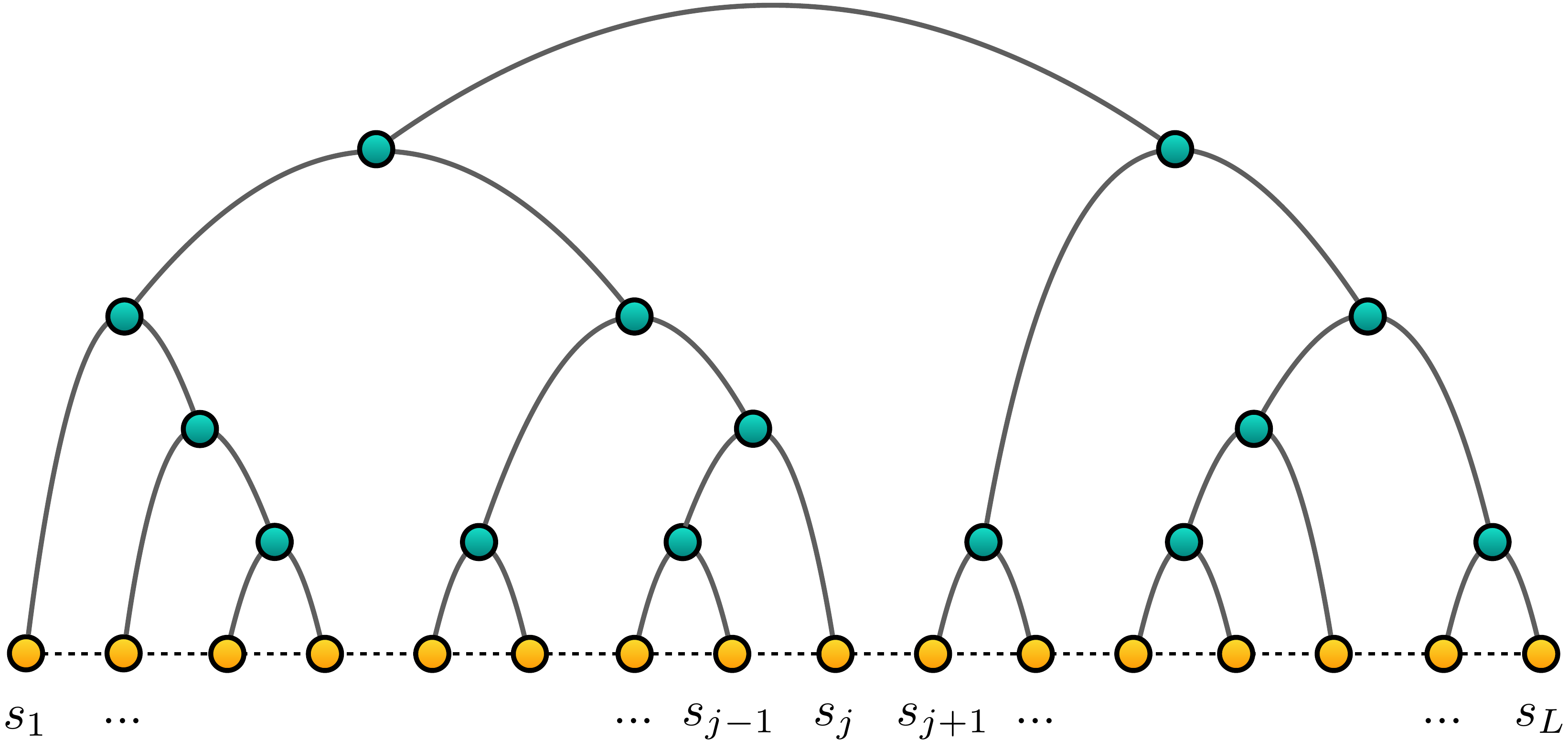}  
  \label{fig:adaptive_tree}
  \newline
\end{subfigure}
\caption{(a) Representation of a homogeneous binary TTN for a one-dimensional spin chain. Yellow circles indicate the sites of the chain connected to the tree by the physical indices; the green circles are the tensors of the TTN. (b) Representation of a TTN completely self-assembled from the nearest-neighbor interactions between the spins of the chain: the strength of the couplings determines the structure of the tree.}
\label{fig:figure1}
\end{figure}

Disordered systems represent still today one of the most challenging problems in condensed matter and statistical physics. The study of low-energy properties of disordered and frustrated magnetic systems, characterized by the presence of quenched, i.e., time-independent, random interactions, is crucial for understanding their phase diagrams, their excitation, and the out-of-equilibrium relaxation dynamics \cite{Vojta2019}.
However, due to the presence of randomness, it is difficult to obtain analytical solutions in the different regimes of parameters of the models and for the great variety of microscopical interactions. Thus, numerical methods play a significant role in determining the ground states and the low-energy properties of disordered systems \cite{Edwards_1972, RevModPhys.44.127, Nozadze_2013, PhysRevB.99.205113}. In particular, when dealing with some degree of randomness, it is crucial to collect extensive statistics over a large number of physical realizations of the disorder and for large systems sizes. In order to get reliable results that approximate the continuum limit properties of the system, these steps could largely increase the memory and the computational time required for the numerical simulations. In this framework, the introduction and development of innovative algorithms capable of efficiently addressing the problem of disordered quantum systems play a key role in studying their physical properties.    

In the last decades, Tensor Network (TN) algorithms have proved remarkably successful in simulating the static and dynamical properties of quantum many-body systems \cite{doi:10.1080/14789940801912366, PhysRevB.94.165116, Montangero_book, ORUS2014117, Or_s_2019}. These computational techniques benefit from an efficient representation of the low-energy states of the quantum systems, by exploiting area-law bounds for the entanglement and by compressing the relevant information of the exponentially large wave functions into a network of tensors interconnected through auxiliary indices with bond dimension $m$ \cite{RevModPhys.82.277, book_quantum_info}. Also, they do not suffer from the infamous sign problem of Quantum Monte Carlo methods, allowing the study of frustrated and fermionic systems even in the regimes of finite density \cite{PhysRevB.41.9301, Ba_uls_2017}. The main representations of quantum many-body states in terms of TN ansatze are Matrix Product States (MPS) for 1D systems \cite{MPS_1, MPS_2, MPS_3} , Projected Entangled Pair States (PEPS) for 2D systems \cite{PhysRevLett.96.220601, 2004cond.mat.7066V, tepaske2020threedimensional}, Multiscale Entanglement Renormalization Ansatz (MERA)  \cite{PhysRevLett.99.220405, PhysRevLett.102.180406} and Tree Tensor Networks (TTN) that can be defined for both one- and higher-dimensional systems \cite{PhysRevA.74.022320, PhysRevB.90.125154, PhysRevA.81.062335, TN_Anthology, aTTN_2021, qian2021tree, magnifico2020lattice} . A representation of a homogeneous binary TTN for a one-dimensional spin chain is shown in Fig \ref{fig:binary_tree}. 

TNs have been successfully applied in the simulation of disordered systems, especially exploiting the tensor-network strong-disorder renormalization group (tSDRG) \cite{PhysRevB.60.12116, PhysRevB.89.214203}. This powerful algorithm was proposed as a renormalization group numerical method for the one-dimensional Heisenberg model with random interactions and then applied to a variety of spin chains \cite{PhysRevB.96.064427, Tsai_2020}. Recently, it has been successfully applied to two-dimensional random quantum spin systems even in the regime of strong-randomness \cite{Seki_2020, seki2021entanglementbased}.


It exists a close relationship between the tSDRG algorithm and the Tree Tensor Networks (TTN) \cite{PhysRevB.89.214203}. Indeed, in the tSDRG, 
the system is organized in blocks composed of the original spins, then the strongest link connecting two blocks is determined, and the resulting pair of blocks is numerically renormalized into a new upper block with a truncated basis. The procedure is iterated from bottom to top, leading to a tree representation of the ground-state of the system, as schematized for a one-dimensional spin chain in Fig. \ref{fig:adaptive_tree}. In this case, the structure of the network depends on the interactions between the nearest-neighbor spins. This property led to the idea of directly using adaptive and self-assembling binary TTNs for the numerical study of disordered spin chains \cite{PhysRevB.89.214203}. Let us notice that in this case, the structure of the final TTN can be completely ``unbalanced", meaning that there could be very small regions of spins not connected to the intermediate layers and the distance between two lattice sites, in terms of links connecting them, could scale very differently from the optimal logarithmic scaling offered by the homogeneous binary TTNs (Fig. \ref{fig:binary_tree}) \cite{TN_Anthology, Cataldi_2021}, which are the most advantageous tree structures for non-disordered quantum many-body systems. 

In this work, we explore the possibility of achieving a balance between the capabilities of adaptive and self-assembling trees and the properties of homogeneous networks to study of disordered systems. In particular, we propose an adaptive binary TTNs, by introducing a {\it weight parameter} that allows interpolating between the standard homogeneous binary TTNs (Fig. \ref{fig:binary_tree}) and the TTNs completely self-assembled from the nearest-neighbor interactions (Fig. \ref{fig:adaptive_tree}). In order to test the performance obtained with the resulting {\it adaptive-weighted} TTN ansatz, we focus on the ground state properties of the two-dimensional disordered Ising model in a transverse field at zero temperature, showing a clear improvement of numerical precision as a function of the weight parameter, especially for large system.

The article is organized as follows. In Sec. \ref{sec:disordered_ising} we introduce the two-dimensional quantum Ising model in the presence of random disorder. In Sec. \ref{sec:entanglement_law} we study the entanglement scaling of the system for different realisations of the disorder by means of homogeneous binary TTNs. In Sec. \ref{sec:adaptive_TTN} we present in details the adaptive-weighted TTN ansatz, while Sec. \ref{sec:num_results} is devoted to the presentation of the numerical results that we obtained by using our TTN algorithms. In Sec. \ref{sec:conclusion} we draw our conclusions. Additional details related to the numerical simulations are presented in the Appendix.

\section{Disordered Ising model in transverse field}\label{sec:disordered_ising}
In order to test the capabilities of our {\it adaptive-weighted} TTN ansatz, we consider the 2D quantum Ising model in a uniform transverse magnetic field along the $z$-direction and in presence of quenched random disorder and frustration \cite{dutta2015quantum, PhysRevLett.72.4141, Rieger}. For a $L \times L$ square lattice, the Hamiltonian is 
\begin{equation}\label{eq:Hamiltonian}
H = -\sum_{\left < j,k \right >} \lambda_{j k} \sigma^x_j \sigma^x_k - h\sum_l \sigma^z_l
\end{equation}
where $\left < j,k \right>$ indicates the sum over the nearest-neighbor sites of the lattice. The operator $\sigma^\alpha_j$ refers to the standard Pauli matrix in the $\alpha$-direction, with $\alpha=x,y,z$, defined on the site $j$ and satisfying the commutations relations $\left [ \sigma^\alpha_j, \sigma^\beta_k \right ] = 2i \delta_{j k} \epsilon_{\alpha \beta \gamma} \sigma^\gamma_j$, where $\delta_{jk}$ is the Kronecker delta and $\epsilon_{\alpha, \beta, \gamma}$ is the Levi-Civita symbol. The external magnetic field $h$ is homogeneous, while the exchange interactions $\lambda_{j k}$ are site-dependent uncorrelated random variables. In particular, we keep $h$ fixed and we generate the realizations of the disorder with the $\left \{ \lambda_{j k} \right \}_{j k}$ being normally and independently distributed with zero-mean and unit-variance. 

At zero temperature, the model described by Eq. \eqref{eq:Hamiltonian} exhibits a second order phase transition,  driven by quantum fluctuations as a function of the value of $h$. It exists a critical value $h_c$ separating a spin-glass ordered phase for $h < h_c$ from a disordered paramagnetic phase for $h > h_c$, with $h_c \approx 0.608$ \cite{PhysRevLett.72.4141}.

\section{Entanglement distribution: area law}\label{sec:entanglement_law}


\begin{figure}
\begin{subfigure}{.5\textwidth}
\caption{}
  \centering
  \includegraphics[width=.9\linewidth]{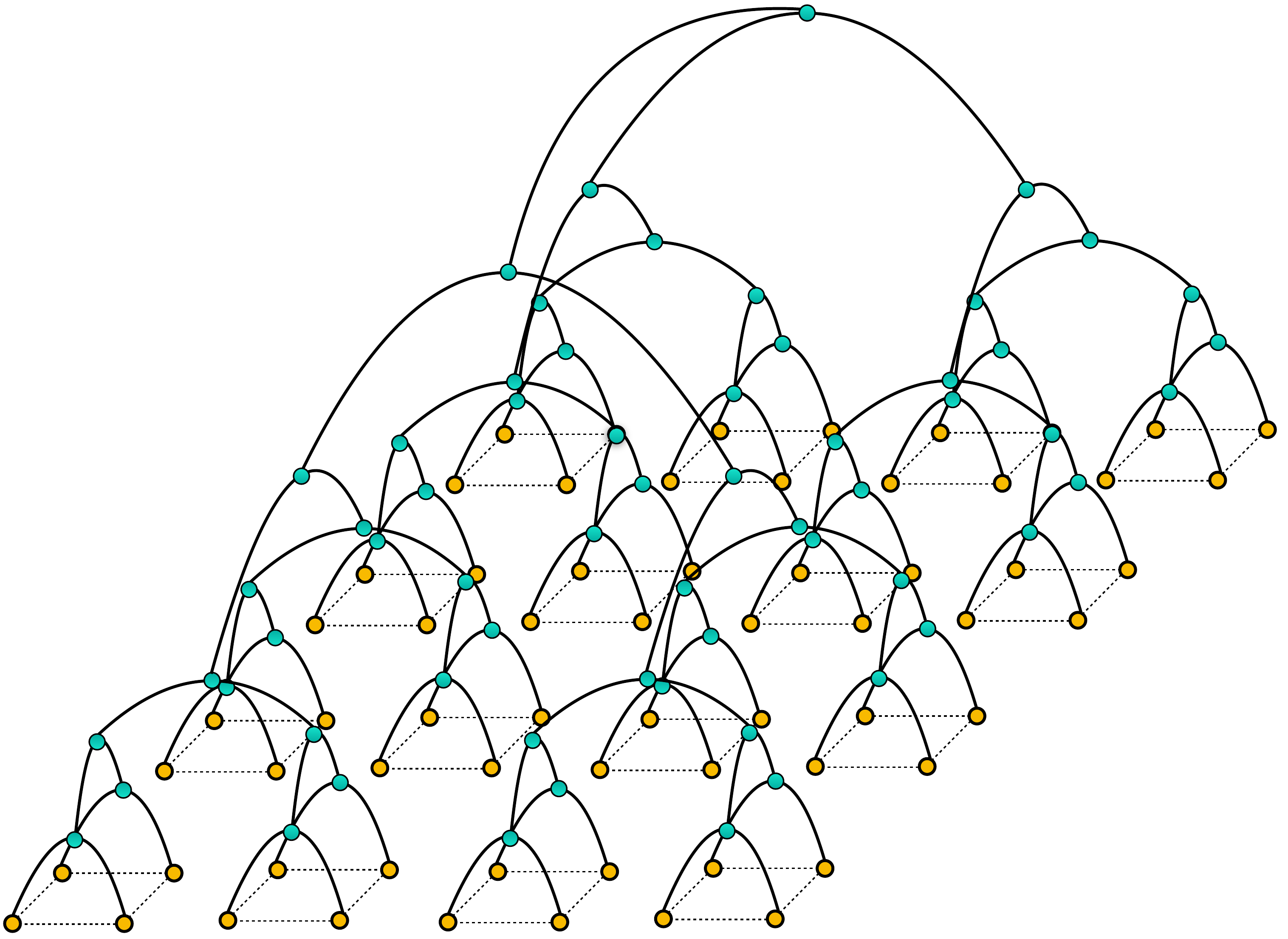}  
  \label{fig:2D_standard_TTN}
  \newline
\end{subfigure}
\newline
\begin{subfigure}{.5\textwidth}
  \caption{}
  \centering
  \includegraphics[width=.65\linewidth]{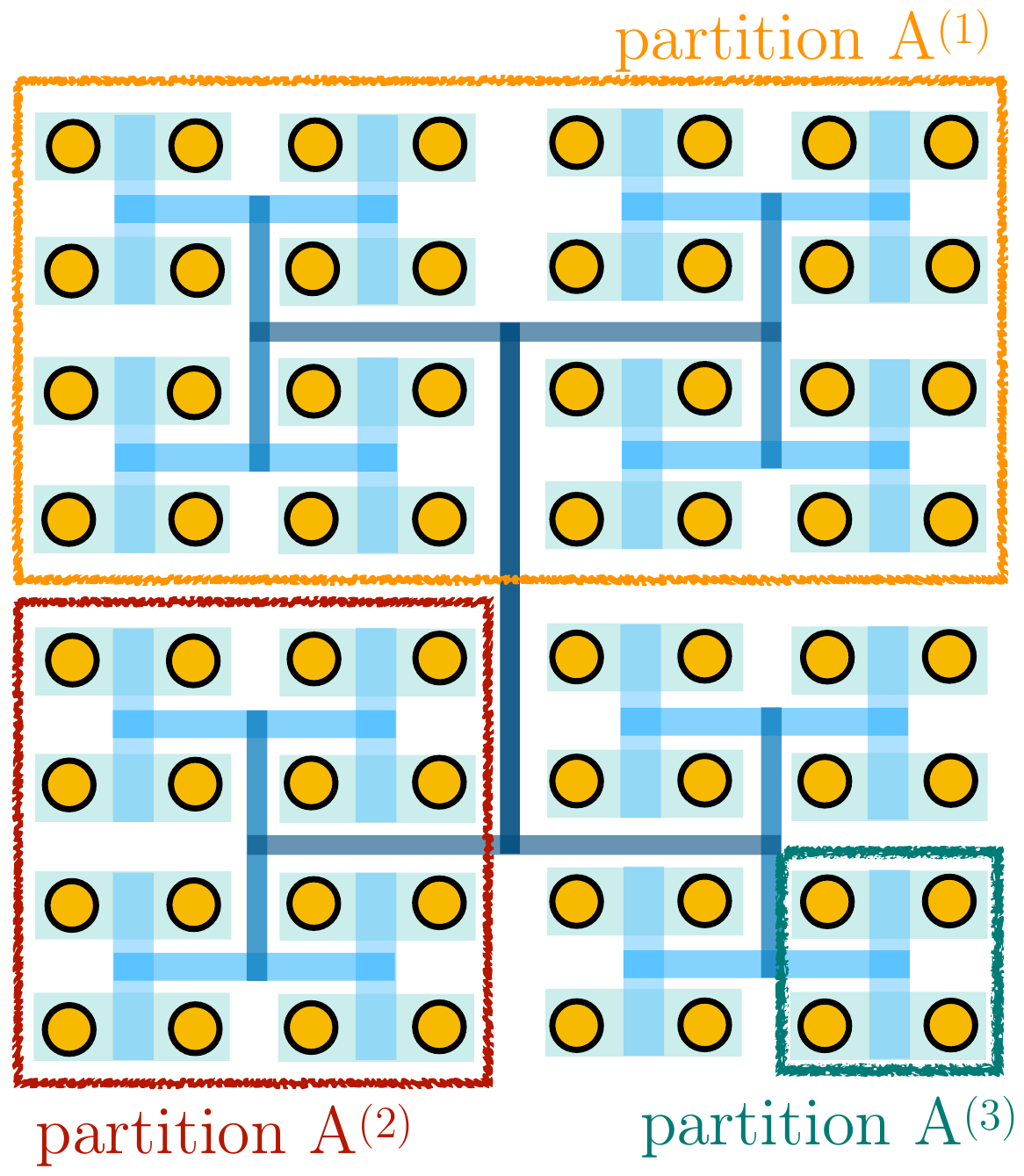}  
  \label{fig:projection_TTN}
  \newline
\end{subfigure}
\caption{(a) Two-dimensional homogeneous TTN ansatz. Yellow circles indicate the physical sites of the lattice connected to the tree by physical indices; green circles are the tensors making up the TTN. (b) Projection of the TTN structure on the plane of the physical lattice. The three partitions used for the computation of the entanglement entropy are highlighted.}
\label{fig:figure2_both}
\end{figure}

At the heart of the tensor network algorithms and their capabilities to efficiently approximate ground states of locally interacting Hamiltonians is the {\it area law} scaling for the entanglement entropy \cite{RevModPhys.82.277}. The area law implies that, if we consider a partition of the system, being the latter in its ground state, the entanglement entropy scales linearly with the boundary area of the region (with possible small logarithmic corrections). Thus, the amount of the entanglement and quantum correlations in the ground states of locally interacting Hamiltonians is much smaller with respect to a generic quantum many-body state (for instance, a highly excited state or a thermal state), for which the entanglement entropy scales extensively with the volume of the considered partition. This property allows for an efficient representation, in terms of tensor networks, of the low-energy states of locally interacting Hamiltonians \cite{Montangero_book, book_quantum_info}.

It has been found that area law for entanglement entropy is valid for a large class of quantum many-body systems, both for one-dimensional and higher-dimensional Hamiltonians \cite{area_law_2016}. 

The entanglement entropy can be quantified by using the standard von-Neumann entropy \cite{nielsen_chuang_2010}. Let us consider a system defined on a lattice, prepared in the ground state $\left | \psi \right >$, and a generic partition $A$ of the lattice and its complement $B$. Starting from the density matrix of the full system $\rho = \left | \psi \right > \left < \psi \right |$, it is possible to obtain the reduced density matrix of the partition $A$ by tracing out the degrees of freedom of $B$, i.e. $\rho_A = \mathrm{Tr}_B \rho = \mathrm{Tr}_B \left | \psi \right > \left < \psi \right | $. The von-Neumann entanglement entropy related to the partition $A$ is defined as 

\begin{equation}\label{eq:entropy_Def}
S(\rho_A) = -\mathrm{Tr}(\rho_A \mathrm{log_2} \rho_A).
\end{equation}
This entropy quantifies the amount of entanglement and quantum correlations that the partition $A$ shares with its complement $B$. In this language, the area law for locally interacting Hamiltonians implies that 

\begin{equation}\label{eq:area_law}
S(\rho_A) \sim  \left | \partial A \right |
\end{equation}
where $\left | \partial A \right |$ is the boundary area of the partition $A$ \cite{RevModPhys.82.277}.

In the presence of random disorder, area law is not always satisfied. Indeed, there are many examples of area law violations in the context of disordered and inhomogeneous systems \cite{area_law_2016, area_law_violation_2019}. However, for the specific case of the random Ising model in Eq. \eqref{eq:Hamiltonian}, it has been numerically found that the leading term of the entanglement entropy always scales linearly with the area of the partition \cite{disorder_ising_area_law_2008, disorder_ising_area_law_2012}.


\begin{figure}
\begin{subfigure}{0.49\textwidth}
\caption{}
  \centering
  \includegraphics[width=0.985\textwidth]{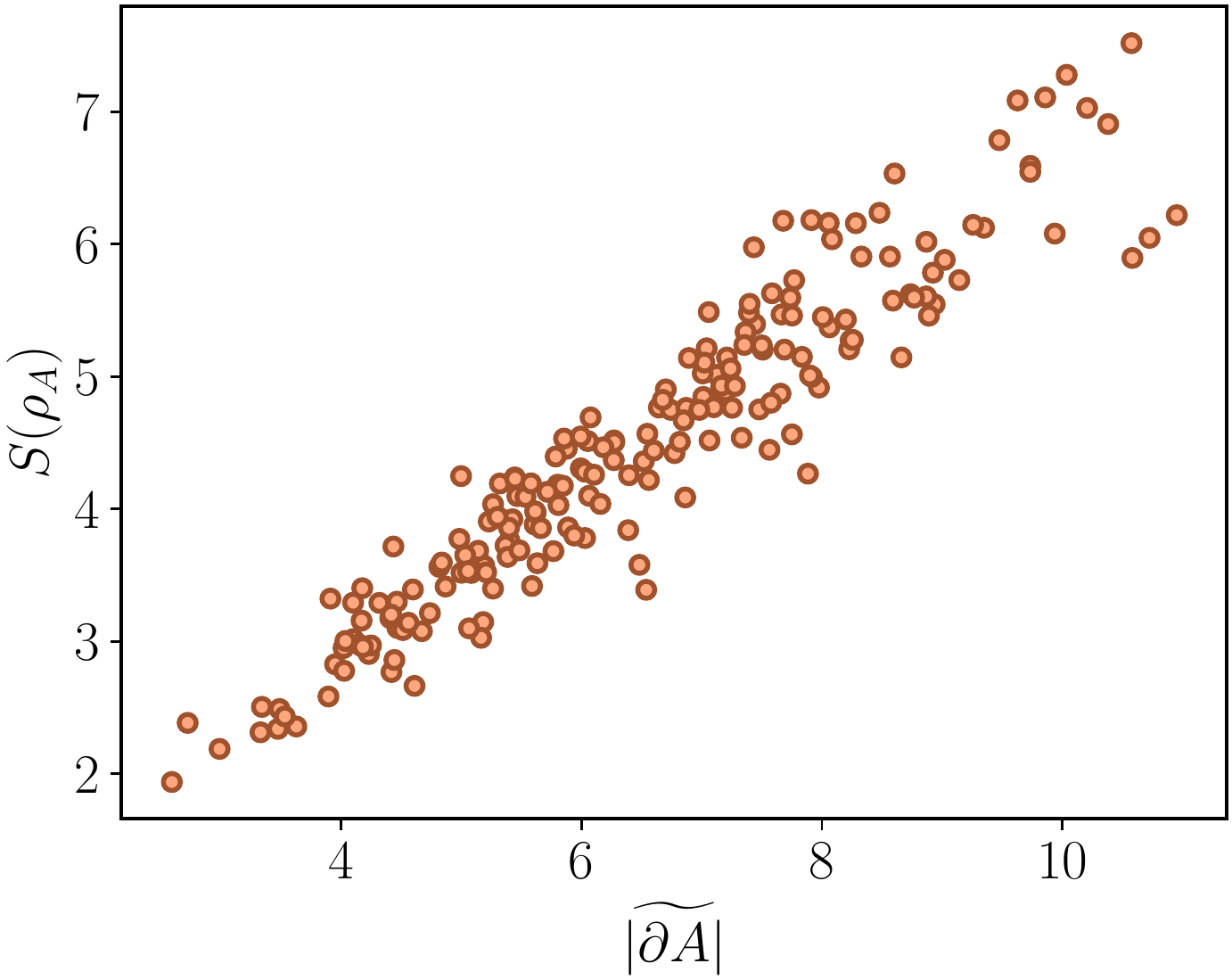}  
  \label{fig:cut_1}
\end{subfigure}
\par\medskip
\begin{subfigure}{.245\textwidth}
  \caption{}
  \centering
  \includegraphics[width=.95\textwidth]{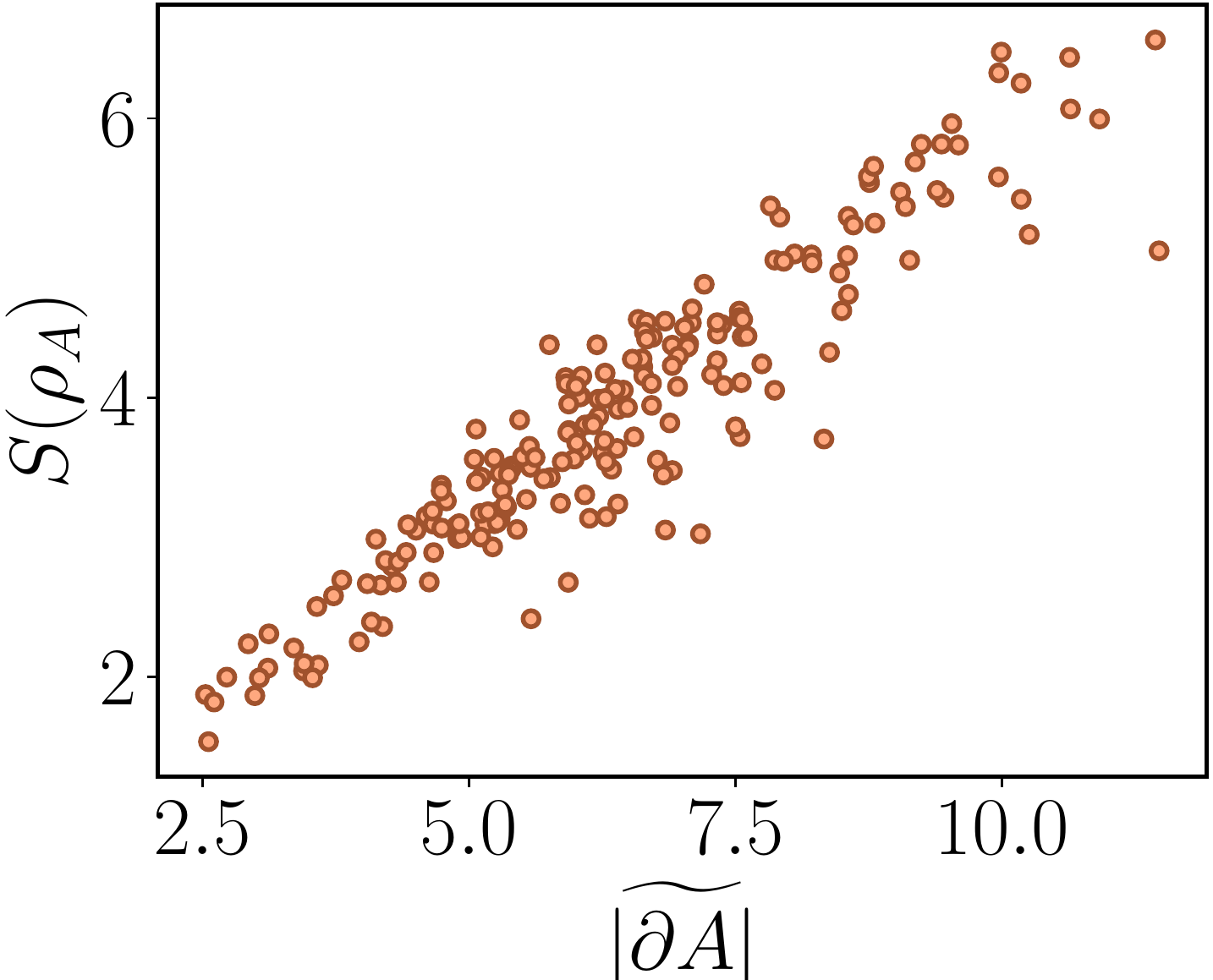}  
  \label{fig:cut_2}
\end{subfigure}%
\begin{subfigure}{.245\textwidth}
\caption{}
  \centering
  \includegraphics[width=.95\textwidth]{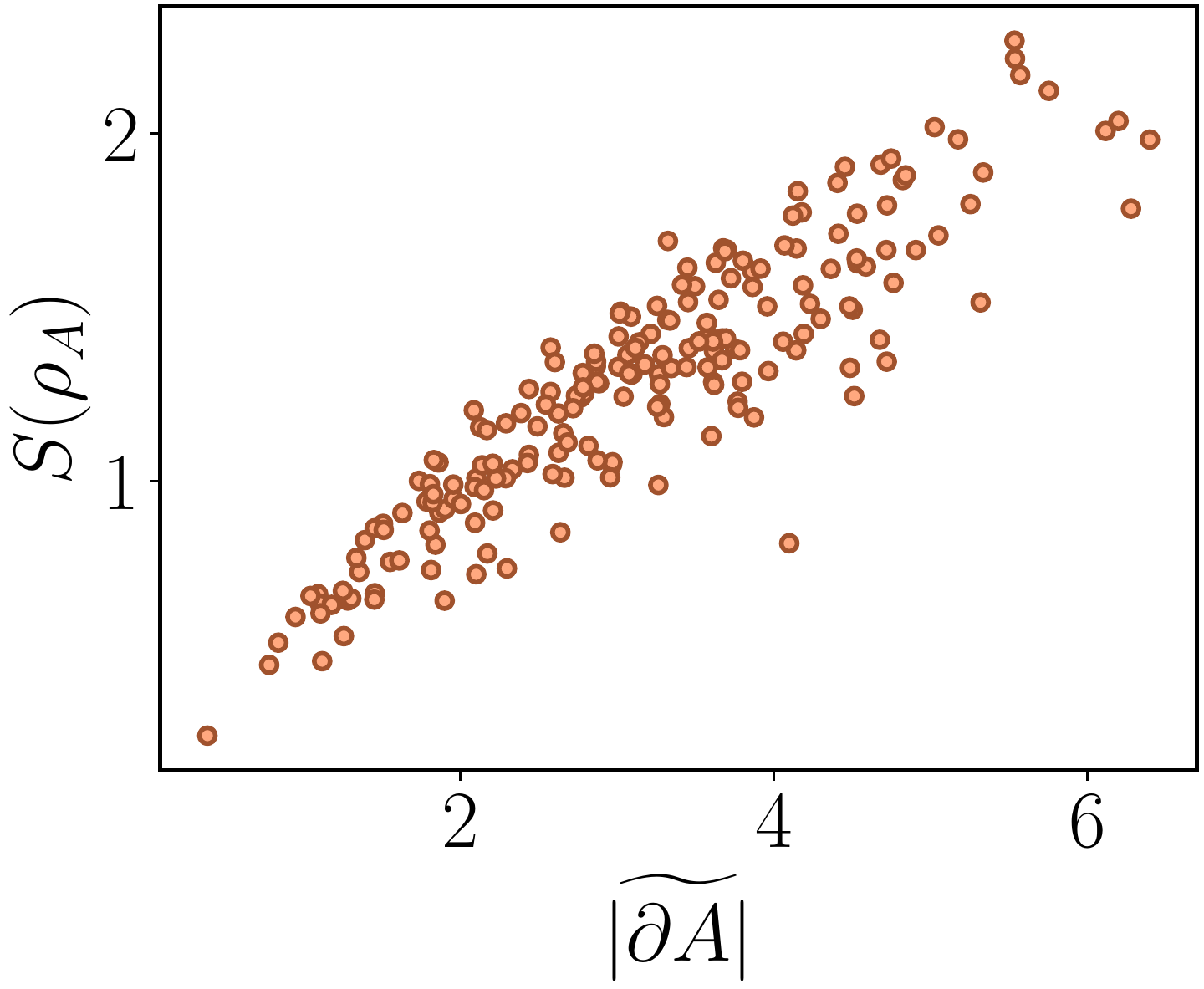}  
  \label{fig:cut_3}
  \newline
\end{subfigure}
\caption{Behavior of the von-Neumann entanglement entropy as a function of the effective area of (a) the partition $A^{(1)}$, (b) partition $A^{(2)}$, and (c) partition $A^{(3)}$. We consider the case with $L=8$ and 200 realisations of the random disorder.}
\label{fig:entropy_data}
\end{figure}

In order to construct our {\it adaptive-weighted} TTN ansatz, we are interested in the scaling of the entanglement entropy as a function of the strength of the random couplings of Eq. \eqref{eq:Hamiltonian}. Indeed, it is reasonable to expect that the strength of the couplings must enter the relation in Eq. \eqref{eq:area_law} in some way: if, as an extreme scenario, all the interactions along the boundary of considered partition $A$ happen to be null, we expect no correlations (and hence no entanglement) to build up between the two subsystems $A$ and its complement $B$. By continuity, we also expect weak couplings to produce small correlations. For these reasons, we define an effective area for the generic partition $A$, i.e. the area of the partition weighted with the strength of the couplings crossing the boundaries: 
\begin{equation}
\widetilde{ \left |  \partial A \right | } := \sum_{(j,k) \in \partial A } \left | \lambda_{jk} \right |
\end{equation}

\begin{figure*}[t]
    \includegraphics[width=0.99\textwidth]{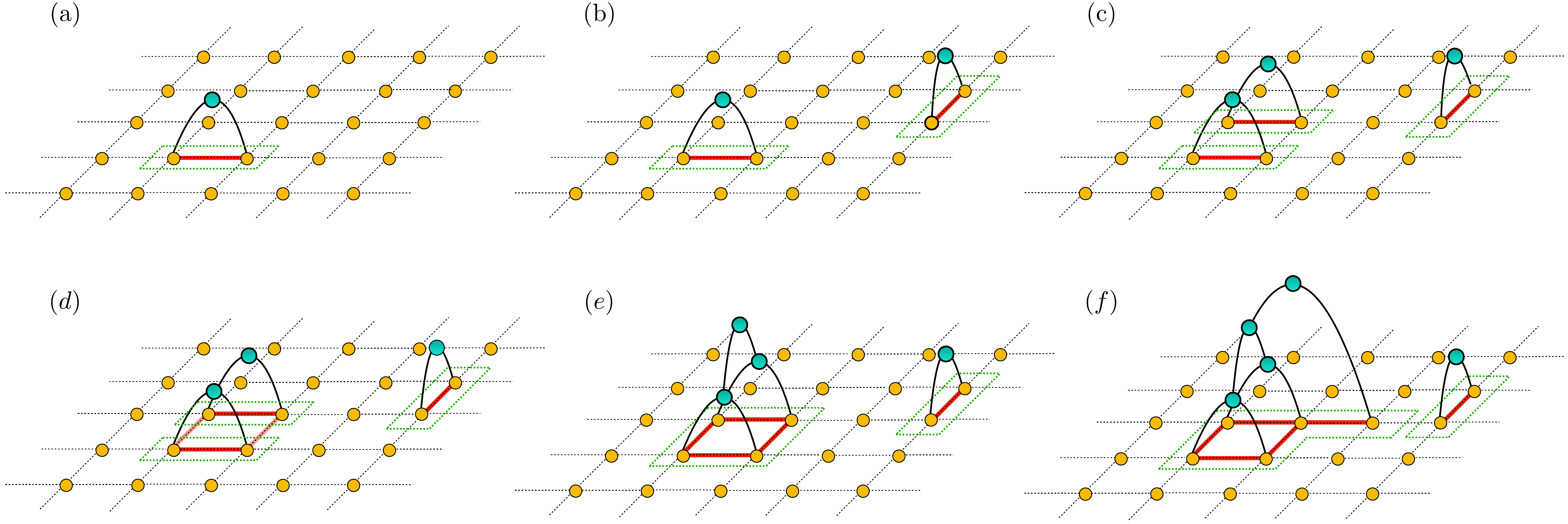}
    \newline
    \caption{Construction of the adaptive-weighted TTN for two-dimensional lattice and a specific realisation of the random disorder. (a) A tensor is placed on top of two spins that interact with the strongest coupling in absolute value: the two original spins are then considered as a single effective spin (dashed green line). (b) This procedure is iterated for the next strongest coupling and (c) so forth. (d) If two effective spins share a border larger than one unit, the couplings between them are summed in absolute value. (e) If the newly defined coupling is the strongest one at this stage, a tensor is placed on the two underlying effective spins. (f) The procedure is iterated for the remaining couplings of the lattice.
    In all the previous steps, the considered couplings for the construction of the adaptive-weighted TTN depend on a weight parameter $\alpha$ as defined in Eq. \eqref{eq:def_couplings}.}
    \label{fig:tree_adaptive_scheme}
\end{figure*}

Taking into account the  Hamiltonian in Eq. \eqref{eq:Hamiltonian} with $h=2$, we generate 200 realizations of the random disorder, and we numerically determine the ground state of the system at zero temperature by using homogeneous binary TTN for two-dimensional lattices (see Fig. \ref{fig:figure2_both}). For more details about the algorithms and their implementation, see Appendix \ref{appendix:num_simulations}.  

We consider three partitions, $A^{(1)}$, $A^{(2)}$, and $A^{(3)}$, as shown in Fig. \ref{fig:projection_TTN}. Partition $A^{(1)}$ is defined by the cut of the highest link of the tree and corresponds to one-half of the lattice; partition $A^{(2)}$ involves the cuts of two upper-links of the tree and corresponds to one-fourth of the lattice, whereas partition $A^{(3)}$ corresponds to one-sixteenth of the lattice. For each of these partitions we compute the von-Neumann entanglement entropy by using the definition in  Eq. \eqref{eq:entropy_Def}. The results for the different realizations of the disorder, i.e., for different values of the effective area $\widetilde{ \left | \partial A \right |}$, are reported in Fig. \ref{fig:entropy_data}. A linear dependence between the two quantities is clearly shown for the three partitions. Thus, the amount of the entanglement contained in the ground state of the system is directly proportional to the effective area of the considered partition and, consequently, to the strength of the crossed couplings. This property will serve as a starting point for the construction of the {\it adaptive-weighted} TTN ansatz presented in the next section.

\section{Adaptive-weighted Tree Tensor Networks}\label{sec:adaptive_TTN}

The linear scaling of the entanglement entropy as a function of the effective area is a helpful property that we want to exploit to improve the numerical precision of TTN simulations of disordered and inhomogeneous systems. 

The main idea is to start building the TTN from the most entangled groups of spins. Indeed, the more we ``climb up" the tree, the bigger the subsystems connected by the links of the TTN become. The bigger the subsystems, the higher the entanglement carried by the links becomes. Given that the amount of entanglement we can store to keep the simulation manageable is fixed by the bond dimension of the links $m$, the description of the quantum correlations becomes less and less faithful as we go from the bottom to the top of the tree. So, it is reasonable to connect with the uppermost internal links of the TTN the least entangled pairs of subsystems.

In the previous section, we observed a linear dependence relating the entanglement between two subsystems and the total strength of the couplings between them, i.e., the effective area $\widetilde{ \left |  \partial A \right | }$. Since the couplings are known a priori (contrarily to entanglement, which requires the knowledge of the ground state), we can exploit this relation and start building an optimized TTN from the most interacting subsystems. 

The algorithm that we implement for each realization of the disorder of the Hamiltonian in Eq. \eqref{eq:Hamiltonian} is described in the following steps:

\begin{enumerate}
    \item We start by identifying the strongest (in absolute value) coupling between two spins
of the whole system. We attach a link to each one of these two spins, and we connect them
with a tensor (see Fig. \ref{fig:tree_adaptive_scheme}(a)). The two spins connected to the tensor are now defined to be a single
effective spin (represented by the dashed green line in Fig. \ref{fig:tree_adaptive_scheme}), interacting with its neighbors. The number of physical spins
forming an effective spin is called its ``volume".
   \item We repeat the procedure for the next strongest couplings, treating the effective spins on the same footing as the
others, as shown in Figs. \ref{fig:tree_adaptive_scheme}(b) and \ref{fig:tree_adaptive_scheme}(c).
   \item In case two effective spins share a border with an area larger than one (see Fig.  \ref{fig:tree_adaptive_scheme}(d)), the effective coupling between them is taken to be the sum of the absolute values of the original couplings. The effective coupling is then treated on the same footing as the others. This renormalization procedure for the couplings is chosen in agreement with the results of the previous section: the renormalized coupling between two effective spins is precisely the effective area they share, and hence is related to the entanglement between them.
   \item At this stage, if the strongest coupling is the newly redefined one, a tensor is placed on it, as shown in Fig. \ref{fig:tree_adaptive_scheme}(e).
   \item The procedure is iterated for the remaining couplings on the lattice: a possible step is shown, for instance, in Fig. \ref{fig:tree_adaptive_scheme}(f).
\end{enumerate}

We stress the fact that the renormalized couplings introduced here are used only to construct the tree, whereas in the ground state search algorithm only the original Hamiltonian is considered. 

This protocol allows maximizing the effective area associated with the internal links of the lowest levels of the TTN and, in turn, minimizing the one associated with the uppermost internal links, where the approximation introduced by the bond dimension is most relevant. Note that this procedure resembles the one presented in \cite{PhysRevB.89.214203}, where adaptive TTNs were used for the first time for disordered one-dimensional systems. However, our approach for two-dimensional TTNs, is directly based on the distribution of the random couplings in the Hamiltonian instead of the distributions of the local energy gaps, and hence it does not require additional diagonalizations. 

Furthermore,  we introduce a key element in our protocol, a tunable {\it weight parameter} that allows to interpolate between the homogeneous binary TTN (Fig. \ref{fig:2D_standard_TTN}) and the TTN completely self-assembled from the nearest-neighbor random interactions. In particular, if $V(\sigma)$ is the volume of the effective spin $\sigma$, we applied steps (1)-(5) by considering $\widetilde{\lambda^{\alpha}_{j,k}}$ in place of the original couplings $\lambda_{jk}$ of Eq. \eqref{eq:Hamiltonian}, where
\begin{equation}\label{eq:def_couplings}
\widetilde{\lambda^{\alpha}_{jk}} = \frac{\lambda_{jk}}{\left ( V(\sigma_j) + V(\sigma_k) \right )^\alpha}.
\end{equation}
As a result, the weight parameter $\alpha \geq 0 $ favors connecting effective spins with small volumes first. In such a way, it is less likely to arrive at higher levels of the TTN with spins with small volumes not yet connected, hence ending up connecting very small volume spins to very big volume ones. Thus, the weight parameter $\alpha$ favors binary-like geometries during the construction of the optimized TTN: for $\alpha=0$ the tree is completely self-assembled from the couplings, whereas, in the limit of very large-$\alpha$, the tree tends to be more and more similar to the standard homogeneous binary TTN of Fig. \ref{fig:2D_standard_TTN}.

In the following section, we see that this parameter affects the numerical precision of the simulations in a non-trivial way, and the best performances are obtained for intermediate values of $\alpha$, where a balance between the descriptive power of the completely self-assembled TTNs and the standard homogeneous binary TTNs is reached. 

\section{Numerical Results}\label{sec:num_results}

\begin{figure}
\begin{subfigure}{0.45\textwidth}
\caption{}
  \centering
  \includegraphics[width=.95\textwidth]{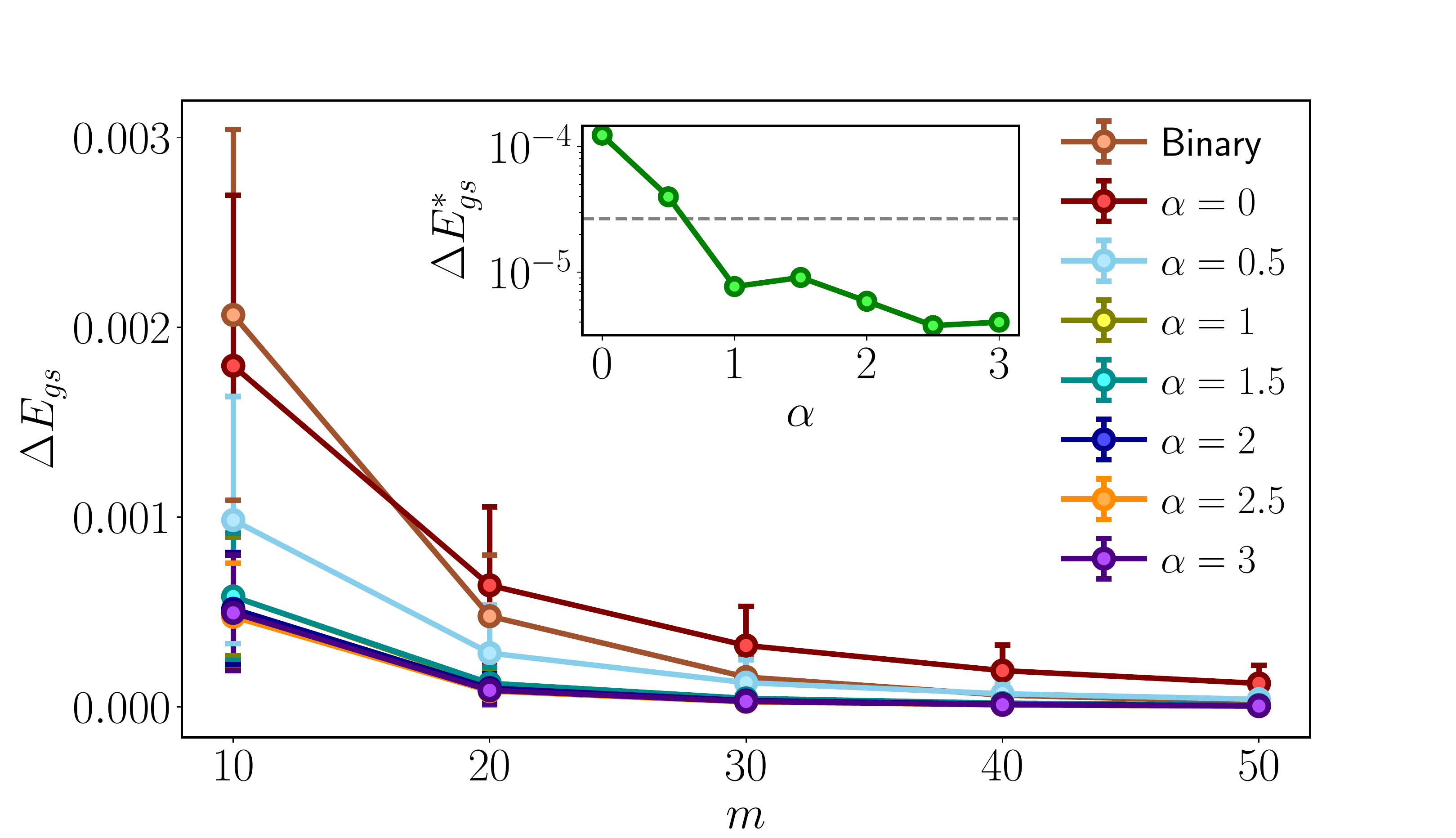}  
  \label{fig:results_1}
\end{subfigure}
\newline
\begin{subfigure}{.45\textwidth}
  \caption{}
  \centering
  \includegraphics[width=.95\textwidth]{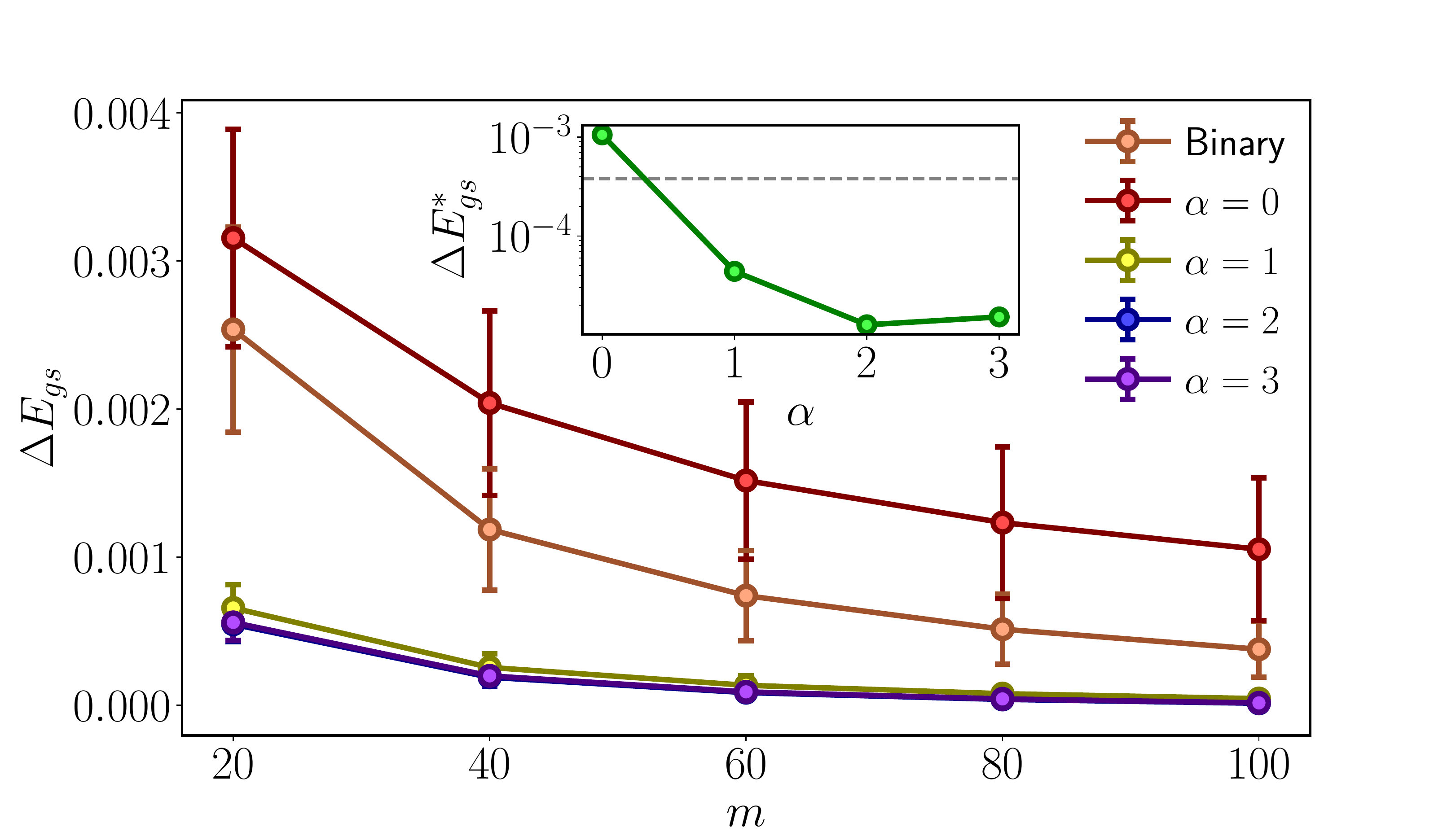}  
  \label{fig:results_2}
\end{subfigure}
\newline
\begin{subfigure}{.45\textwidth}
\caption{}
  \centering
  \includegraphics[width=.95\textwidth]{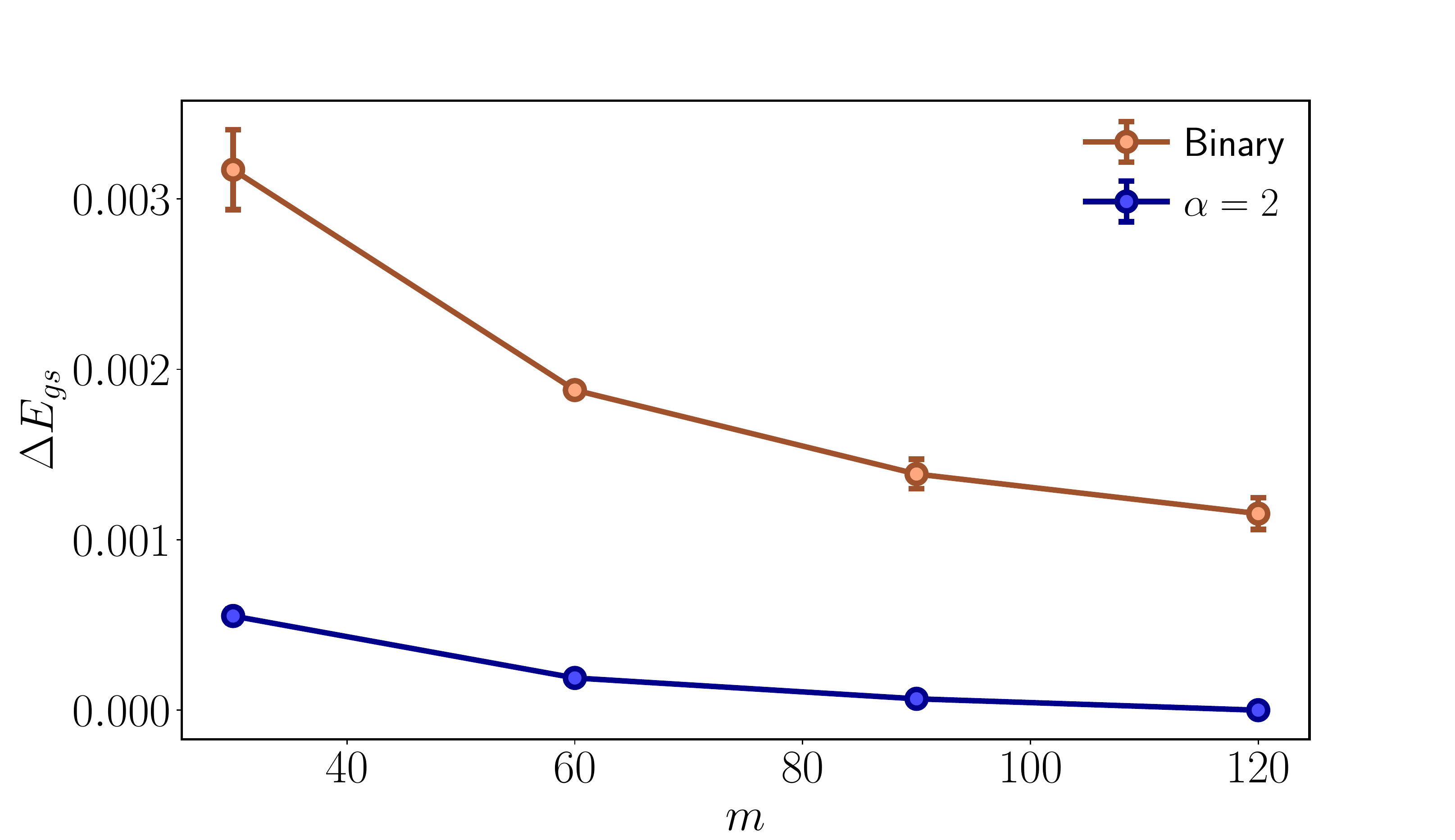}  
  \label{fig:results_3}
  \newline
\end{subfigure}
\caption{Average ground state energy for $h=2$, for different values of the weight parameter, as a function of the bond dimension $m$, and (a) $L=8$, (b) $L=16$, (c) $L=32$. Error bars refer to the standard deviation of the average over the different realisations of the disorder. Insets in (a) and (b):  average energy, in logarithmic scale, obtained for the largest bond dimension as a function of $\alpha$; the dashed line represents the energy obtained by using the homogeneous binary TTN.}
\label{fig:num_results}
\end{figure}

\begin{figure}[t]
\begin{subfigure}{0.45\textwidth}
    \caption{}
    \centering
    \includegraphics[width=0.96\textwidth]{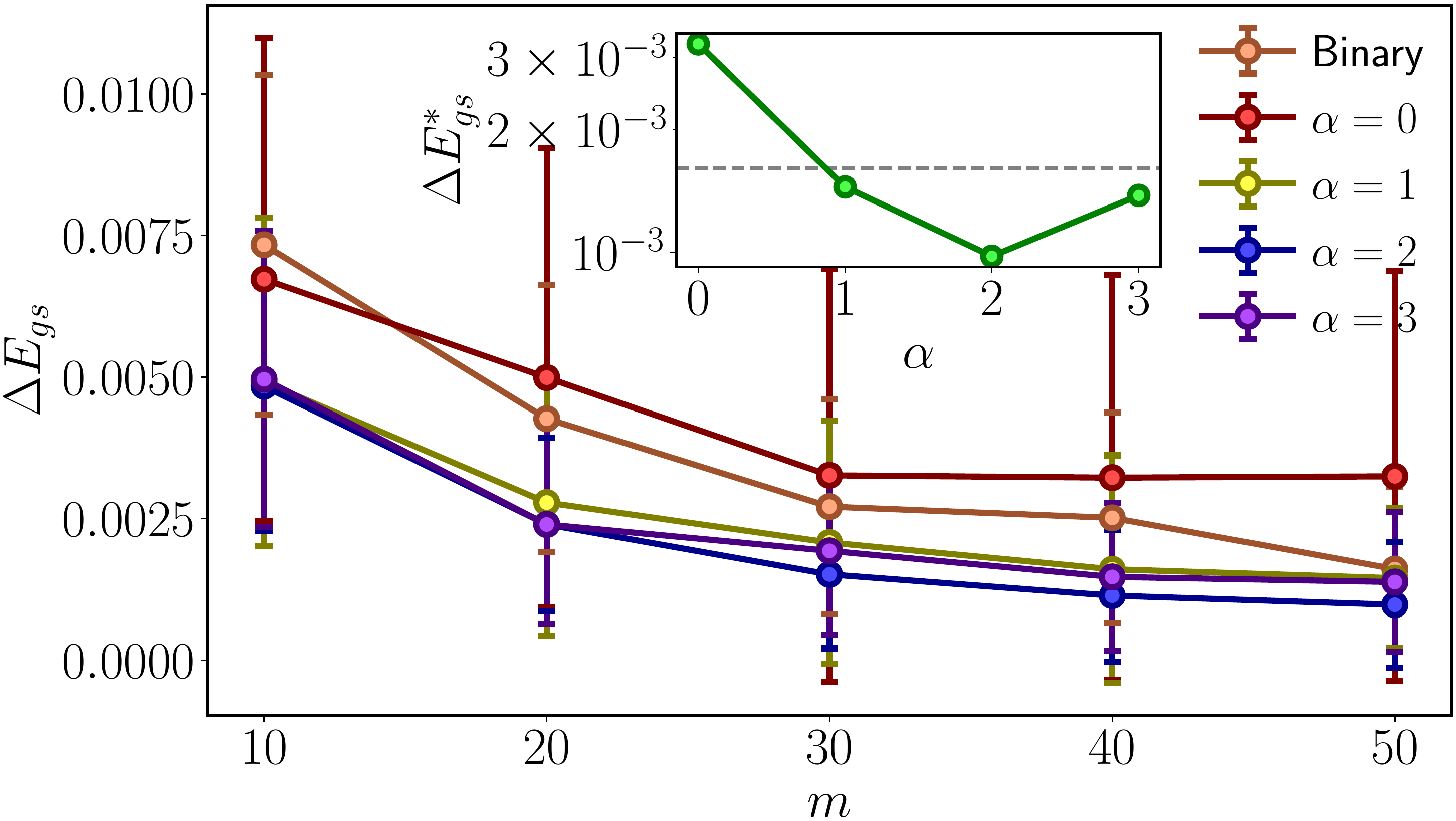}
    \label{fig:results_critical_point}
\end{subfigure}
\newline
\begin{subfigure}{0.45\textwidth}
    \caption{}
    \centering
    \includegraphics[width=0.95\textwidth]{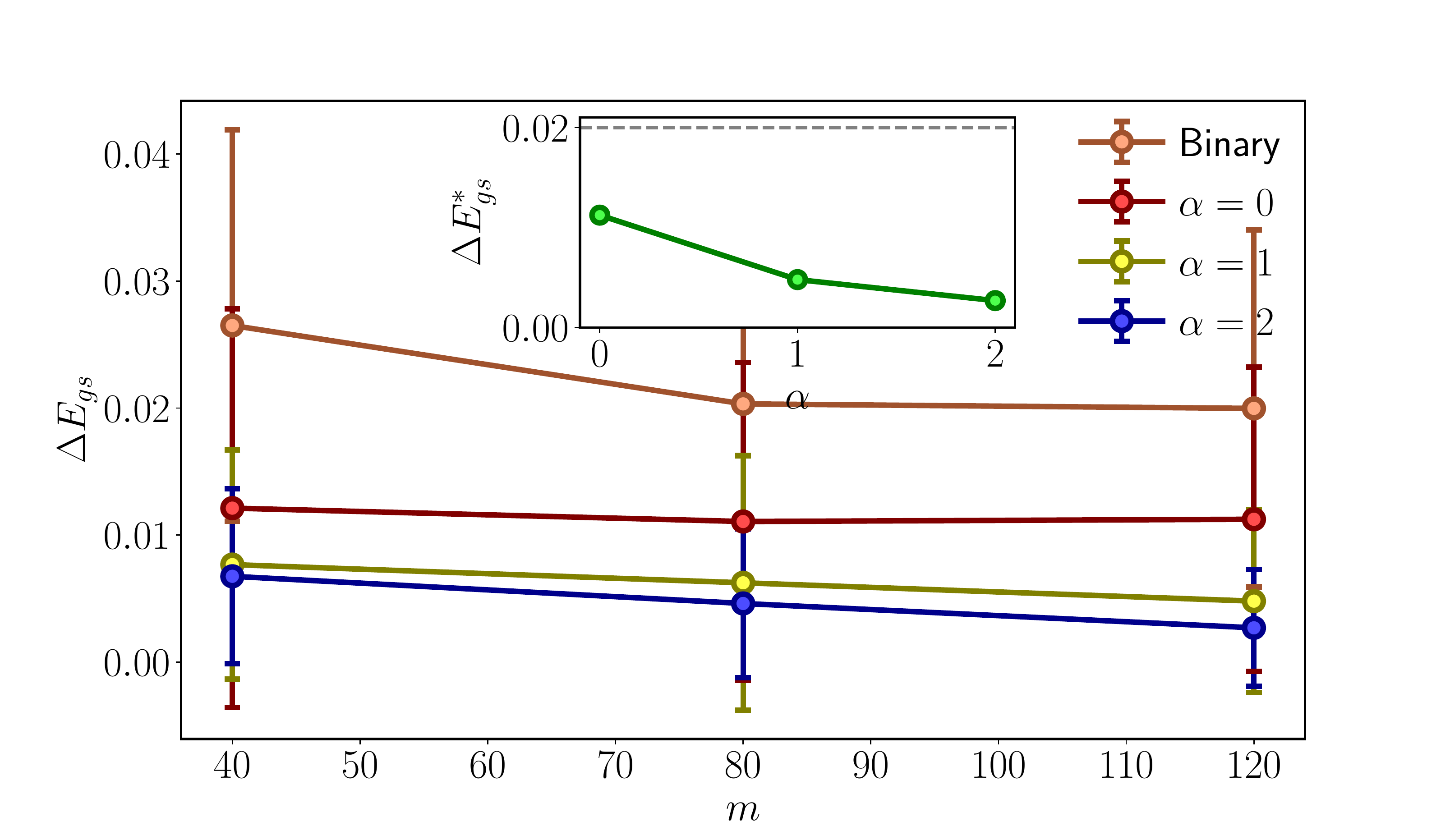}
    \label{fig:results_small_field}
    \newline
\end{subfigure}
\caption{Average ground state energy for (a) $h=0.61$ (close to the critical point) and (b) $h=0.0$ (spin-glass phase), for different values of the weight parameter, as a function of the bond dimension $m$, and $L=8$. Error bars refer to the standard deviation of the average over the different realisations of the disorder. Insets in (a) and (b): average energy obtained for the largest bond dimension as a function of $\alpha$; the dashed line represents the energy obtained by using the homogeneous binary TTN.}
\label{}
\end{figure}

Hereafter we present the results for the ground state energy of the Hamiltonian in Eq. \eqref{eq:Hamiltonian} obtained by using the adaptive-weighted TTNs for several values of $\alpha$, and the standard homogeneous binary TTN. In particular, we generated independent realisations of the disorder, and for each of them, we applied the ground state searching algorithm for different values of the bond dimension $m$ (for more details, see Appendix \ref{appendix:num_simulations}). 

We stress that the limit of $\alpha=0$ of our protocol is not strictly equivalent to the algorithm presented in \cite{PhysRevB.89.214203}, since the construction of the tree and the optimization of the tree tensors are not carried out simultaneously, but in two different steps.

For comparability purposes, all the energies $E_i$ of the different realisations of the disorder are rescaled to the lowest value $E_0$ obtained in all the simulations by computing the quantity $\Delta E_{i} = (E_i - E_0)/E_0$. Then, the average over the different realisations is taken.

In Fig. \ref{fig:num_results}, we show the results of the average ground state energy of the Hamiltonian in Eq. \eqref{eq:Hamiltonian} with $h=2$ (disordered paramagnetic phase), as a function of the bond dimension, for different values of the weight parameter $\alpha$ and for different lattice sizes up to $32 \times 32$. The number of independent realisations of the disorder that we generated depends on the size, i.e., 200 realizations for $L=8$, 50 realisations for $L=16$ and 6 realisations for $L=32$. In all plots, we also report the error bars referring to the standard deviation of the average over the disorder. We observe that the ground state average energy decreases as a function of the bond dimension, converging to an asymptotic value. Also, we notice that the standard homogeneous binary TTNs outperform the completely self-assembled TTNs (the cases with $\alpha=0$). However, a clear and major improvement for any value of the bond dimension is achieved when the adaptive-weighted TTNs with $\alpha \ge 1$ are adopted, especially for large system sizes. Thus, the weight parameter $\alpha$ allows constructing adaptive and optimized TTNs that significantly improve the numerical precision of the simulations, also in the regime of small to intermediate values of the bond dimension. 

It is worth noting that the case with $\alpha=0.5$ for $L=8$, reported in Fig. \ref{fig:results_1}, outperforms the standard homogeneous binary TTNs for small bond dimensions, whereas they both offer similar performances in the regime of large $m$. 

The effect of the weight parameter is particularly evident in the insets of Figs. \ref{fig:results_1}-\ref{fig:results_2}, where we plot, in logarithmic scale, the average energy obtained for the largest bond dimension as a function of $\alpha$, and we represent with the dashed line the energy obtained by using the homogeneous binary TTN (i.e. the limit of $\alpha \to \infty$). The best performances are obtained for intermediate values of $\alpha \approx 2$, being the average energy higher for $\alpha=3$ and in the limit of the homogeneous binary TTN.

We also tested the performance of our protocol when simulating the system close to the critical point. In Fig. \ref{fig:results_critical_point} we report the results of the average ground state energy for $h=0.61$, $L=8$ and 300 realisations of the disorder. Also in this case, the adaptive-weighted TTNs outperform the binary and the completely self-assembled TTNs, with the best performances obtained for $\alpha = 2$.

Analogous results are observed in the spin-glass ordered phase of the Hamiltonian in Eq. \eqref{eq:Hamiltonian}, as shown in in Fig. \ref{fig:results_small_field} for $h=0.0$, $L=8$ and 50 realisations of the disorder. In this case, the adaptive-weighted TTNs outperform the binary TTNs for any value of $\alpha$, returning lower energy for all the values of the bond dimension. 

All these results, for the disorder paramagnetic phase, the critical regime, and the spin-glass phase, show that the adaptive-weighted TTNs benefit from an optimal balance between the capabilities of the self-assembling TTNs and the ones of the standard binary TTNs, representing a valuable and efficient ansatz for the study of disordered and inhomogeneous quantum many-body systems.

\section{Conclusions}\label{sec:conclusion}
In this work, we have examined the application of TTN variational algorithms to the numerical study of two-dimensional quantum many-body systems in the presence of random disorder. In particular, we have proposed a protocol that, by starting from the knowledge of the Hamiltonian of the system, allows the construction of a new ansatz, the adaptive-weighted TTNs. This structure is assembled, for each realisation of the disorder, taking into account the strength of the nearest-neighbor interactions and thus intrinsically reflecting the distribution of the entanglement into the system. Furthermore, the presence of a weight parameter prevents the TTN from being completely unbalanced, ensuring a better convergence during the variational optimization of the ground state searching process.  

In order to test the performances of the adaptive-weighted TTNs, we have considered the two-dimensional quantum Ising model in the presence of random disorder, and we have carried out a comparison between the ground state average energies obtained for different values of the weight-parameter, with a particular focus on the two cases of the standard homogeneous TTNs and the completely self-assembled TTNs. Our results show that the adaptive-weighted TTNs  offer a clear advantage with respect to the other TTN structures, leading to an improvement in the numerical precision for any value of the bond dimension of the network. 

In conclusion, the adaptive-weighted TTN ansatz introduced in this work provides an efficient and powerful tool for numerically simulating disordered and inhomogeneous quantum many-body systems. Also, our protocol could be generalized to higher-dimensional lattices or applied to the study of strong inhomogeneous systems, such as atoms and molecules, that are relevant in the context of quantum chemistry \cite{doi:10.1063/1.4798639}.

\section*{Acknowledgments}
This work is partially supported by the Italian PRIN2017 and Fondazione CARIPARO, the INFN project QUANTUM, the QuantERA projects QTFLAG and QuantHEP, the Horizon 2020 research and innovation programme under grant agreement No 817482 (Quantum Flagship - PASQuanS). We acknowledge computational resources by the Cloud Veneto, CINECA, the BwUniCluster, and by University of Padova Strategic Research Infrastructure Grant 2017: ``CAPRI: Calcolo ad Alte Prestazioni per la Ricerca e l'Innovazione".
\newline

\appendix

\section{Numerical simulations}\label{appendix:num_simulations}
The numerical simulations presented in this work have been performed by using the Tree Tensor Networks variational ground-state searching algorithm \cite{TN_Anthology}.
Given the Hamiltonian for each realisation of the disorder, the homogenoeus or the optimised TTN with a fixed bond dimension $m$ is constructed and the ground state is determined by optimising each of the tensor in the network. This procedure is carried out by exploiting the Krylov sub-space expansion in order to derive a local eigenvalue problem for each tensor \cite{TN_Anthology}, which is numerically solved by applying the Arnoldi method of the ARPACK library \cite{Lehoucq97arpackusers}. This step is iterated for all the tensors in the TTN by starting from the lowest layers. The whole procedure is repeated several times, iteratively reducing the expectation value of the energy. For each value of the bond dimension, the convergence of the final energy is below $10^{-6}$.


\begin{thebibliography}{0}%
\makeatletter
\providecommand \@ifxundefined [1]{%
 \@ifx{#1\undefined}
}%
\providecommand \@ifnum [1]{%
 \ifnum #1\expandafter \@firstoftwo
 \else \expandafter \@secondoftwo
 \fi
}%
\providecommand \@ifx [1]{%
 \ifx #1\expandafter \@firstoftwo
 \else \expandafter \@secondoftwo
 \fi
}%
\providecommand \natexlab [1]{#1}%
\providecommand \enquote  [1]{``#1''}%
\providecommand \bibnamefont  [1]{#1}%
\providecommand \bibfnamefont [1]{#1}%
\providecommand \citenamefont [1]{#1}%
\providecommand \href@noop [0]{\@secondoftwo}%
\providecommand \href [0]{\begingroup \@sanitize@url \@href}%
\providecommand \@href[1]{\@@startlink{#1}\@@href}%
\providecommand \@@href[1]{\endgroup#1\@@endlink}%
\providecommand \@sanitize@url [0]{\catcode `\\12\catcode `\$12\catcode
  `\&12\catcode `\#12\catcode `\^12\catcode `\_12\catcode `\%12\relax}%
\providecommand \@@startlink[1]{}%
\providecommand \@@endlink[0]{}%
\providecommand \url  [0]{\begingroup\@sanitize@url \@url }%
\providecommand \@url [1]{\endgroup\@href {#1}{\urlprefix }}%
\providecommand \urlprefix  [0]{URL }%
\providecommand \Eprint [0]{\href }%
\providecommand \doibase [0]{http://dx.doi.org/}%
\providecommand \selectlanguage [0]{\@gobble}%
\providecommand \bibinfo  [0]{\@secondoftwo}%
\providecommand \bibfield  [0]{\@secondoftwo}%
\providecommand \translation [1]{[#1]}%
\providecommand \BibitemOpen [0]{}%
\providecommand \bibitemStop [0]{}%
\providecommand \bibitemNoStop [0]{.\EOS\space}%
\providecommand \EOS [0]{\spacefactor3000\relax}%
\providecommand \BibitemShut  [1]{\csname bibitem#1\endcsname}%
\let\auto@bib@innerbib\@empty
\end{thebibliography}%


\begin{thebibliography}{62}%
\makeatletter
\providecommand \@ifxundefined [1]{%
 \@ifx{#1\undefined}
}%
\providecommand \@ifnum [1]{%
 \ifnum #1\expandafter \@firstoftwo
 \else \expandafter \@secondoftwo
 \fi
}%
\providecommand \@ifx [1]{%
 \ifx #1\expandafter \@firstoftwo
 \else \expandafter \@secondoftwo
 \fi
}%
\providecommand \natexlab [1]{#1}%
\providecommand \enquote  [1]{``#1''}%
\providecommand \bibnamefont  [1]{#1}%
\providecommand \bibfnamefont [1]{#1}%
\providecommand \citenamefont [1]{#1}%
\providecommand \href@noop [0]{\@secondoftwo}%
\providecommand \href [0]{\begingroup \@sanitize@url \@href}%
\providecommand \@href[1]{\@@startlink{#1}\@@href}%
\providecommand \@@href[1]{\endgroup#1\@@endlink}%
\providecommand \@sanitize@url [0]{\catcode `\\12\catcode `\$12\catcode
  `\&12\catcode `\#12\catcode `\^12\catcode `\_12\catcode `\%12\relax}%
\providecommand \@@startlink[1]{}%
\providecommand \@@endlink[0]{}%
\providecommand \url  [0]{\begingroup\@sanitize@url \@url }%
\providecommand \@url [1]{\endgroup\@href {#1}{\urlprefix }}%
\providecommand \urlprefix  [0]{URL }%
\providecommand \Eprint [0]{\href }%
\providecommand \doibase [0]{http://dx.doi.org/}%
\providecommand \selectlanguage [0]{\@gobble}%
\providecommand \bibinfo  [0]{\@secondoftwo}%
\providecommand \bibfield  [0]{\@secondoftwo}%
\providecommand \translation [1]{[#1]}%
\providecommand \BibitemOpen [0]{}%
\providecommand \bibitemStop [0]{}%
\providecommand \bibitemNoStop [0]{.\EOS\space}%
\providecommand \EOS [0]{\spacefactor3000\relax}%
\providecommand \BibitemShut  [1]{\csname bibitem#1\endcsname}%
\let\auto@bib@innerbib\@empty
\bibitem [{\citenamefont {Vojta}(2019)}]{Vojta2019}%
  \BibitemOpen
  \bibfield  {author} {\bibinfo {author} {\bibfnamefont {T.}~\bibnamefont
  {Vojta}},\ }\href {\doibase 10.1146/annurev-conmatphys-031218-013433}
  {\bibfield  {journal} {\bibinfo  {journal} {Annual Review of Condensed Matter
  Physics}\ }\textbf {\bibinfo {volume} {10}},\ \bibinfo {pages} {233}
  (\bibinfo {year} {2019})}\BibitemShut {NoStop}%
\bibitem [{\citenamefont {Vojta}(2006)}]{Vojta_2006}%
  \BibitemOpen
  \bibfield  {author} {\bibinfo {author} {\bibfnamefont {T.}~\bibnamefont
  {Vojta}},\ }\href {\doibase 10.1088/0305-4470/39/22/r01} {\bibfield
  {journal} {\bibinfo  {journal} {Journal of Physics A: Mathematical and
  General}\ }\textbf {\bibinfo {volume} {39}},\ \bibinfo {pages} {R143}
  (\bibinfo {year} {2006})}\BibitemShut {NoStop}%
\bibitem [{\citenamefont {Vojta}(2010)}]{Vojta_2010}%
  \BibitemOpen
  \bibfield  {author} {\bibinfo {author} {\bibfnamefont {T.}~\bibnamefont
  {Vojta}},\ }\href {\doibase 10.1007/s10909-010-0205-4} {\bibfield  {journal}
  {\bibinfo  {journal} {Journal of Low Temperature Physics}\ }\textbf {\bibinfo
  {volume} {161}},\ \bibinfo {pages} {299} (\bibinfo {year}
  {2010})}\BibitemShut {NoStop}%
\bibitem [{\citenamefont {Rieger}\ and\ \citenamefont {Young}(2007)}]{Rieger}%
  \BibitemOpen
  \bibfield  {author} {\bibinfo {author} {\bibfnamefont {H.}~\bibnamefont
  {Rieger}}\ and\ \bibinfo {author} {\bibfnamefont {A.~P.}\ \bibnamefont
  {Young}},\ }\href {\doibase 10.1007/bfb0104832} {\bibfield  {journal}
  {\bibinfo  {journal} {Lecture Notes in Physics}\ ,\ \bibinfo {pages} {256}}
  (\bibinfo {year} {2007})}\BibitemShut {NoStop}%
\bibitem [{\citenamefont {{Edwards}}\ and\ \citenamefont
  {{Anderson}}(1975)}]{Spin_Glass_1}%
  \BibitemOpen
  \bibfield  {author} {\bibinfo {author} {\bibfnamefont {S.~F.}\ \bibnamefont
  {{Edwards}}}\ and\ \bibinfo {author} {\bibfnamefont {P.~W.}\ \bibnamefont
  {{Anderson}}},\ }\href {\doibase 10.1088/0305-4608/5/5/017} {\bibfield
  {journal} {\bibinfo  {journal} {Journal of Physics F Metal Physics}\ }\textbf
  {\bibinfo {volume} {5}},\ \bibinfo {pages} {965} (\bibinfo {year}
  {1975})}\BibitemShut {NoStop}%
\bibitem [{\citenamefont {Mezard}\ \emph {et~al.}(1986)\citenamefont {Mezard},
  \citenamefont {Parisi},\ and\ \citenamefont {Virasoro}}]{Spin_Glass_2}%
  \BibitemOpen
  \bibfield  {author} {\bibinfo {author} {\bibfnamefont {M.}~\bibnamefont
  {Mezard}}, \bibinfo {author} {\bibfnamefont {G.}~\bibnamefont {Parisi}}, \
  and\ \bibinfo {author} {\bibfnamefont {M.}~\bibnamefont {Virasoro}},\ }\href
  {\doibase 10.1142/0271} {\emph {\bibinfo {title} {Spin Glass Theory and
  Beyond}}}\ (\bibinfo  {publisher} {World Scientific},\ \bibinfo {year}
  {1986})\BibitemShut {NoStop}%
\bibitem [{\citenamefont {Mossi}\ \emph {et~al.}(2017)\citenamefont {Mossi},
  \citenamefont {Parolini}, \citenamefont {Pilati},\ and\ \citenamefont
  {Scardicchio}}]{Mossi_2017}%
  \BibitemOpen
  \bibfield  {author} {\bibinfo {author} {\bibfnamefont {G.}~\bibnamefont
  {Mossi}}, \bibinfo {author} {\bibfnamefont {T.}~\bibnamefont {Parolini}},
  \bibinfo {author} {\bibfnamefont {S.}~\bibnamefont {Pilati}}, \ and\ \bibinfo
  {author} {\bibfnamefont {A.}~\bibnamefont {Scardicchio}},\ }\href {\doibase
  10.1088/1742-5468/aa5286} {\bibfield  {journal} {\bibinfo  {journal} {Journal
  of Statistical Mechanics: Theory and Experiment}\ }\textbf {\bibinfo {volume}
  {2017}},\ \bibinfo {pages} {013102} (\bibinfo {year} {2017})}\BibitemShut
  {NoStop}%
\bibitem [{\citenamefont {Anderson}(1958)}]{PhysRev.109.1492}%
  \BibitemOpen
  \bibfield  {author} {\bibinfo {author} {\bibfnamefont {P.~W.}\ \bibnamefont
  {Anderson}},\ }\href {\doibase 10.1103/PhysRev.109.1492} {\bibfield
  {journal} {\bibinfo  {journal} {Phys. Rev.}\ }\textbf {\bibinfo {volume}
  {109}},\ \bibinfo {pages} {1492} (\bibinfo {year} {1958})}\BibitemShut
  {NoStop}%
\bibitem [{\citenamefont {Abrahams}(2010)}]{doi:10.1142/7663}%
  \BibitemOpen
  \bibfield  {author} {\bibinfo {author} {\bibfnamefont {E.}~\bibnamefont
  {Abrahams}},\ }\href {\doibase 10.1142/7663} {\emph {\bibinfo {title} {50
  Years of Anderson Localization}}}\ (\bibinfo  {publisher} {World
  Scientific},\ \bibinfo {year} {2010})\BibitemShut {NoStop}%
\bibitem [{\citenamefont {Evers}\ and\ \citenamefont
  {Mirlin}(2008)}]{RevModPhys.80.1355}%
  \BibitemOpen
  \bibfield  {author} {\bibinfo {author} {\bibfnamefont {F.}~\bibnamefont
  {Evers}}\ and\ \bibinfo {author} {\bibfnamefont {A.~D.}\ \bibnamefont
  {Mirlin}},\ }\href {\doibase 10.1103/RevModPhys.80.1355} {\bibfield
  {journal} {\bibinfo  {journal} {Rev. Mod. Phys.}\ }\textbf {\bibinfo {volume}
  {80}},\ \bibinfo {pages} {1355} (\bibinfo {year} {2008})}\BibitemShut
  {NoStop}%
\bibitem [{\citenamefont {Belitz}\ and\ \citenamefont
  {Kirkpatrick}(1994)}]{RevModPhys.66.261}%
  \BibitemOpen
  \bibfield  {author} {\bibinfo {author} {\bibfnamefont {D.}~\bibnamefont
  {Belitz}}\ and\ \bibinfo {author} {\bibfnamefont {T.~R.}\ \bibnamefont
  {Kirkpatrick}},\ }\href {\doibase 10.1103/RevModPhys.66.261} {\bibfield
  {journal} {\bibinfo  {journal} {Rev. Mod. Phys.}\ }\textbf {\bibinfo {volume}
  {66}},\ \bibinfo {pages} {261} (\bibinfo {year} {1994})}\BibitemShut
  {NoStop}%
\bibitem [{\citenamefont {Abanin}\ and\ \citenamefont
  {Papic}(2017)}]{many_body_localization_1}%
  \BibitemOpen
  \bibfield  {author} {\bibinfo {author} {\bibfnamefont {D.~A.}\ \bibnamefont
  {Abanin}}\ and\ \bibinfo {author} {\bibfnamefont {Z.}~\bibnamefont {Papic}},\
  }\href {\doibase https://doi.org/10.1002/andp.201700169} {\bibfield
  {journal} {\bibinfo  {journal} {Annalen der Physik}\ }\textbf {\bibinfo
  {volume} {529}},\ \bibinfo {pages} {1700169} (\bibinfo {year}
  {2017})}\BibitemShut {NoStop}%
\bibitem [{\citenamefont {Abanin}\ \emph {et~al.}(2019)\citenamefont {Abanin},
  \citenamefont {Altman}, \citenamefont {Bloch},\ and\ \citenamefont
  {Serbyn}}]{Abanin_2019}%
  \BibitemOpen
  \bibfield  {author} {\bibinfo {author} {\bibfnamefont {D.~A.}\ \bibnamefont
  {Abanin}}, \bibinfo {author} {\bibfnamefont {E.}~\bibnamefont {Altman}},
  \bibinfo {author} {\bibfnamefont {I.}~\bibnamefont {Bloch}}, \ and\ \bibinfo
  {author} {\bibfnamefont {M.}~\bibnamefont {Serbyn}},\ }\href {\doibase
  10.1103/RevModPhys.91.021001} {\bibfield  {journal} {\bibinfo  {journal}
  {Rev. Mod. Phys.}\ }\textbf {\bibinfo {volume} {91}},\ \bibinfo {pages}
  {021001} (\bibinfo {year} {2019})}\BibitemShut {NoStop}%
\bibitem [{\citenamefont {Luca}\ and\ \citenamefont
  {Scardicchio}(2013)}]{De_Luca_2013}%
  \BibitemOpen
  \bibfield  {author} {\bibinfo {author} {\bibfnamefont {A.~D.}\ \bibnamefont
  {Luca}}\ and\ \bibinfo {author} {\bibfnamefont {A.}~\bibnamefont
  {Scardicchio}},\ }\href {\doibase 10.1209/0295-5075/101/37003} {\bibfield
  {journal} {\bibinfo  {journal} {{EPL} (Europhysics Letters)}\ }\textbf
  {\bibinfo {volume} {101}},\ \bibinfo {pages} {37003} (\bibinfo {year}
  {2013})}\BibitemShut {NoStop}%
\bibitem [{\citenamefont {L\'opez-Sandoval}\ and\ \citenamefont
  {Garcia}(2006)}]{PhysRevB.74.174204}%
  \BibitemOpen
  \bibfield  {author} {\bibinfo {author} {\bibfnamefont {R.}~\bibnamefont
  {L\'opez-Sandoval}}\ and\ \bibinfo {author} {\bibfnamefont {M.~E.}\
  \bibnamefont {Garcia}},\ }\href {\doibase 10.1103/PhysRevB.74.174204}
  {\bibfield  {journal} {\bibinfo  {journal} {Phys. Rev. B}\ }\textbf {\bibinfo
  {volume} {74}},\ \bibinfo {pages} {174204} (\bibinfo {year}
  {2006})}\BibitemShut {NoStop}%
\bibitem [{\citenamefont {Mohdeb}\ \emph {et~al.}(2020)\citenamefont {Mohdeb},
  \citenamefont {Vahedi}, \citenamefont {Moure}, \citenamefont {Roshani},
  \citenamefont {Lee}, \citenamefont {Bhatt}, \citenamefont {Kettemann},\ and\
  \citenamefont {Haas}}]{PhysRevB.102.214201}%
  \BibitemOpen
  \bibfield  {author} {\bibinfo {author} {\bibfnamefont {Y.}~\bibnamefont
  {Mohdeb}}, \bibinfo {author} {\bibfnamefont {J.}~\bibnamefont {Vahedi}},
  \bibinfo {author} {\bibfnamefont {N.}~\bibnamefont {Moure}}, \bibinfo
  {author} {\bibfnamefont {A.}~\bibnamefont {Roshani}}, \bibinfo {author}
  {\bibfnamefont {H.-Y.}\ \bibnamefont {Lee}}, \bibinfo {author} {\bibfnamefont
  {R.~N.}\ \bibnamefont {Bhatt}}, \bibinfo {author} {\bibfnamefont
  {S.}~\bibnamefont {Kettemann}}, \ and\ \bibinfo {author} {\bibfnamefont
  {S.}~\bibnamefont {Haas}},\ }\href {\doibase 10.1103/PhysRevB.102.214201}
  {\bibfield  {journal} {\bibinfo  {journal} {Phys. Rev. B}\ }\textbf {\bibinfo
  {volume} {102}},\ \bibinfo {pages} {214201} (\bibinfo {year}
  {2020})}\BibitemShut {NoStop}%
\bibitem [{\citenamefont {Ghosh}\ and\ \citenamefont
  {Das}(2021)}]{PhysRevB.103.024202}%
  \BibitemOpen
  \bibfield  {author} {\bibinfo {author} {\bibfnamefont {R.}~\bibnamefont
  {Ghosh}}\ and\ \bibinfo {author} {\bibfnamefont {A.}~\bibnamefont {Das}},\
  }\href {\doibase 10.1103/PhysRevB.103.024202} {\bibfield  {journal} {\bibinfo
   {journal} {Phys. Rev. B}\ }\textbf {\bibinfo {volume} {103}},\ \bibinfo
  {pages} {024202} (\bibinfo {year} {2021})}\BibitemShut {NoStop}%
\bibitem [{\citenamefont {Gelin}\ \emph {et~al.}(2021)\citenamefont {Gelin},
  \citenamefont {Velardo},\ and\ \citenamefont
  {Borrelli}}]{disorder_quantum_chem}%
  \BibitemOpen
  \bibfield  {author} {\bibinfo {author} {\bibfnamefont {M.~F.}\ \bibnamefont
  {Gelin}}, \bibinfo {author} {\bibfnamefont {A.}~\bibnamefont {Velardo}}, \
  and\ \bibinfo {author} {\bibfnamefont {R.}~\bibnamefont {Borrelli}},\ }\href
  {\doibase 10.1063/5.0065896} {\bibfield  {journal} {\bibinfo  {journal} {The
  Journal of Chemical Physics}\ }\textbf {\bibinfo {volume} {155}},\ \bibinfo
  {pages} {134102} (\bibinfo {year} {2021})}\BibitemShut {NoStop}%
\bibitem [{\citenamefont {Edwards}\ and\ \citenamefont
  {Thouless}(1972)}]{Edwards_1972}%
  \BibitemOpen
  \bibfield  {author} {\bibinfo {author} {\bibfnamefont {J.~T.}\ \bibnamefont
  {Edwards}}\ and\ \bibinfo {author} {\bibfnamefont {D.~J.}\ \bibnamefont
  {Thouless}},\ }\href {\doibase 10.1088/0022-3719/5/8/007} {\bibfield
  {journal} {\bibinfo  {journal} {Journal of Physics C: Solid State Physics}\
  }\textbf {\bibinfo {volume} {5}},\ \bibinfo {pages} {807} (\bibinfo {year}
  {1972})}\BibitemShut {NoStop}%
\bibitem [{\citenamefont {Dean}(1972)}]{RevModPhys.44.127}%
  \BibitemOpen
  \bibfield  {author} {\bibinfo {author} {\bibfnamefont {P.}~\bibnamefont
  {Dean}},\ }\href {\doibase 10.1103/RevModPhys.44.127} {\bibfield  {journal}
  {\bibinfo  {journal} {Rev. Mod. Phys.}\ }\textbf {\bibinfo {volume} {44}},\
  \bibinfo {pages} {127} (\bibinfo {year} {1972})}\BibitemShut {NoStop}%
\bibitem [{\citenamefont {Nozadze}\ and\ \citenamefont
  {Vojta}(2013)}]{Nozadze_2013}%
  \BibitemOpen
  \bibfield  {author} {\bibinfo {author} {\bibfnamefont {D.}~\bibnamefont
  {Nozadze}}\ and\ \bibinfo {author} {\bibfnamefont {T.}~\bibnamefont
  {Vojta}},\ }\href {\doibase 10.1002/pssb.201350152} {\bibfield  {journal}
  {\bibinfo  {journal} {physica status solidi (b)}\ }\textbf {\bibinfo {volume}
  {251}},\ \bibinfo {pages} {675} (\bibinfo {year} {2013})}\BibitemShut
  {NoStop}%
\bibitem [{\citenamefont {Mulanix}\ \emph {et~al.}(2019)\citenamefont
  {Mulanix}, \citenamefont {Almada},\ and\ \citenamefont
  {Khatami}}]{PhysRevB.99.205113}%
  \BibitemOpen
  \bibfield  {author} {\bibinfo {author} {\bibfnamefont {M.~D.}\ \bibnamefont
  {Mulanix}}, \bibinfo {author} {\bibfnamefont {D.}~\bibnamefont {Almada}}, \
  and\ \bibinfo {author} {\bibfnamefont {E.}~\bibnamefont {Khatami}},\ }\href
  {\doibase 10.1103/PhysRevB.99.205113} {\bibfield  {journal} {\bibinfo
  {journal} {Phys. Rev. B}\ }\textbf {\bibinfo {volume} {99}},\ \bibinfo
  {pages} {205113} (\bibinfo {year} {2019})}\BibitemShut {NoStop}%
\bibitem [{\citenamefont {Verstraete}\ \emph {et~al.}(2008)\citenamefont
  {Verstraete}, \citenamefont {Murg},\ and\ \citenamefont
  {Cirac}}]{doi:10.1080/14789940801912366}%
  \BibitemOpen
  \bibfield  {author} {\bibinfo {author} {\bibfnamefont {F.}~\bibnamefont
  {Verstraete}}, \bibinfo {author} {\bibfnamefont {V.}~\bibnamefont {Murg}}, \
  and\ \bibinfo {author} {\bibfnamefont {J.}~\bibnamefont {Cirac}},\ }\href
  {\doibase 10.1080/14789940801912366} {\bibfield  {journal} {\bibinfo
  {journal} {Advances in Physics}\ }\textbf {\bibinfo {volume} {57}},\ \bibinfo
  {pages} {143} (\bibinfo {year} {2008})}\BibitemShut {NoStop}%
\bibitem [{\citenamefont {Haegeman}\ \emph {et~al.}(2016)\citenamefont
  {Haegeman}, \citenamefont {Lubich}, \citenamefont {Oseledets}, \citenamefont
  {Vandereycken},\ and\ \citenamefont {Verstraete}}]{PhysRevB.94.165116}%
  \BibitemOpen
  \bibfield  {author} {\bibinfo {author} {\bibfnamefont {J.}~\bibnamefont
  {Haegeman}}, \bibinfo {author} {\bibfnamefont {C.}~\bibnamefont {Lubich}},
  \bibinfo {author} {\bibfnamefont {I.}~\bibnamefont {Oseledets}}, \bibinfo
  {author} {\bibfnamefont {B.}~\bibnamefont {Vandereycken}}, \ and\ \bibinfo
  {author} {\bibfnamefont {F.}~\bibnamefont {Verstraete}},\ }\href {\doibase
  10.1103/PhysRevB.94.165116} {\bibfield  {journal} {\bibinfo  {journal} {Phys.
  Rev. B}\ }\textbf {\bibinfo {volume} {94}},\ \bibinfo {pages} {165116}
  (\bibinfo {year} {2016})}\BibitemShut {NoStop}%
\bibitem [{\citenamefont {Montangero}(2018)}]{Montangero_book}%
  \BibitemOpen
  \bibfield  {author} {\bibinfo {author} {\bibfnamefont {S.}~\bibnamefont
  {Montangero}},\ }\href {\doibase https://doi.org/10.1007/978-3-030-01409-4}
  {\emph {\bibinfo {title} {Introduction to Tensor Network Methods}}}\
  (\bibinfo  {publisher} {Springer International Publishing},\ \bibinfo {year}
  {2018})\BibitemShut {NoStop}%
\bibitem [{\citenamefont {Orus}(2014)}]{ORUS2014117}%
  \BibitemOpen
  \bibfield  {author} {\bibinfo {author} {\bibfnamefont {R.}~\bibnamefont
  {Orus}},\ }\href {\doibase https://doi.org/10.1016/j.aop.2014.06.013}
  {\bibfield  {journal} {\bibinfo  {journal} {Annals of Physics}\ }\textbf
  {\bibinfo {volume} {349}},\ \bibinfo {pages} {117} (\bibinfo {year}
  {2014})}\BibitemShut {NoStop}%
\bibitem [{\citenamefont {Orus}(2019)}]{Or_s_2019}%
  \BibitemOpen
  \bibfield  {author} {\bibinfo {author} {\bibfnamefont {R.}~\bibnamefont
  {Orus}},\ }\href {\doibase 10.1038/s42254-019-0086-7} {\bibfield  {journal}
  {\bibinfo  {journal} {Nature Reviews Physics}\ }\textbf {\bibinfo {volume}
  {1}},\ \bibinfo {pages} {538} (\bibinfo {year} {2019})}\BibitemShut {NoStop}%
\bibitem [{\citenamefont {Eisert}\ \emph {et~al.}(2010)\citenamefont {Eisert},
  \citenamefont {Cramer},\ and\ \citenamefont {Plenio}}]{RevModPhys.82.277}%
  \BibitemOpen
  \bibfield  {author} {\bibinfo {author} {\bibfnamefont {J.}~\bibnamefont
  {Eisert}}, \bibinfo {author} {\bibfnamefont {M.}~\bibnamefont {Cramer}}, \
  and\ \bibinfo {author} {\bibfnamefont {M.~B.}\ \bibnamefont {Plenio}},\
  }\href {\doibase 10.1103/RevModPhys.82.277} {\bibfield  {journal} {\bibinfo
  {journal} {Rev. Mod. Phys.}\ }\textbf {\bibinfo {volume} {82}},\ \bibinfo
  {pages} {277} (\bibinfo {year} {2010})}\BibitemShut {NoStop}%
\bibitem [{\citenamefont {Zeng}\ \emph {et~al.}(2019)\citenamefont {Zeng},
  \citenamefont {Chen}, \citenamefont {Zhou},\ and\ \citenamefont
  {Wen}}]{book_quantum_info}%
  \BibitemOpen
  \bibfield  {author} {\bibinfo {author} {\bibfnamefont {B.}~\bibnamefont
  {Zeng}}, \bibinfo {author} {\bibfnamefont {X.}~\bibnamefont {Chen}}, \bibinfo
  {author} {\bibfnamefont {D.-L.}\ \bibnamefont {Zhou}}, \ and\ \bibinfo
  {author} {\bibfnamefont {X.-G.}\ \bibnamefont {Wen}},\ }\href
  {https://doi.org/10.1007/978-1-4939-9084-9} {\emph {\bibinfo {title} {Quantum
  Information Meets Quantum Matter: From Quantum Entanglement to Topological
  Phases of Many-Body Systems}}},\ \bibinfo {edition} {1st}\ ed.,\ Quantum
  Science and Technology\ (\bibinfo  {publisher} {Springer},\ \bibinfo
  {address} {New York, NY},\ \bibinfo {year} {2019})\BibitemShut {NoStop}%
\bibitem [{\citenamefont {Loh}\ \emph {et~al.}(1990)\citenamefont {Loh},
  \citenamefont {Gubernatis}, \citenamefont {Scalettar}, \citenamefont {White},
  \citenamefont {Scalapino},\ and\ \citenamefont {Sugar}}]{PhysRevB.41.9301}%
  \BibitemOpen
  \bibfield  {author} {\bibinfo {author} {\bibfnamefont {E.~Y.}\ \bibnamefont
  {Loh}}, \bibinfo {author} {\bibfnamefont {J.~E.}\ \bibnamefont {Gubernatis}},
  \bibinfo {author} {\bibfnamefont {R.~T.}\ \bibnamefont {Scalettar}}, \bibinfo
  {author} {\bibfnamefont {S.~R.}\ \bibnamefont {White}}, \bibinfo {author}
  {\bibfnamefont {D.~J.}\ \bibnamefont {Scalapino}}, \ and\ \bibinfo {author}
  {\bibfnamefont {R.~L.}\ \bibnamefont {Sugar}},\ }\href {\doibase
  10.1103/PhysRevB.41.9301} {\bibfield  {journal} {\bibinfo  {journal} {Phys.
  Rev. B}\ }\textbf {\bibinfo {volume} {41}},\ \bibinfo {pages} {9301}
  (\bibinfo {year} {1990})}\BibitemShut {NoStop}%
\bibitem [{\citenamefont {Banuls}\ \emph {et~al.}(2017)\citenamefont {Banuls},
  \citenamefont {Cichy}, \citenamefont {Ignacio~Cirac}, \citenamefont {Jansen},
  \citenamefont {Kuhn},\ and\ \citenamefont {Saito}}]{Ba_uls_2017}%
  \BibitemOpen
  \bibfield  {author} {\bibinfo {author} {\bibfnamefont {M.~C.}\ \bibnamefont
  {Banuls}}, \bibinfo {author} {\bibfnamefont {K.}~\bibnamefont {Cichy}},
  \bibinfo {author} {\bibfnamefont {J.}~\bibnamefont {Ignacio~Cirac}}, \bibinfo
  {author} {\bibfnamefont {K.}~\bibnamefont {Jansen}}, \bibinfo {author}
  {\bibfnamefont {S.}~\bibnamefont {Kuhn}}, \ and\ \bibinfo {author}
  {\bibfnamefont {H.}~\bibnamefont {Saito}},\ }\href {\doibase
  10.1051/epjconf/201713704001} {\bibfield  {journal} {\bibinfo  {journal} {EPJ
  Web of Conferences}\ }\textbf {\bibinfo {volume} {137}},\ \bibinfo {pages}
  {04001} (\bibinfo {year} {2017})}\BibitemShut {NoStop}%
\bibitem [{\citenamefont {{Fannes}}\ \emph {et~al.}(1992)\citenamefont
  {{Fannes}}, \citenamefont {{Nachtergaele}},\ and\ \citenamefont
  {{Werner}}}]{MPS_1}%
  \BibitemOpen
  \bibfield  {author} {\bibinfo {author} {\bibfnamefont {M.}~\bibnamefont
  {{Fannes}}}, \bibinfo {author} {\bibfnamefont {B.}~\bibnamefont
  {{Nachtergaele}}}, \ and\ \bibinfo {author} {\bibfnamefont {R.~F.}\
  \bibnamefont {{Werner}}},\ }\href {\doibase
  https://doi.org/10.1007/BF02099178} {\bibfield  {journal} {\bibinfo
  {journal} {Communications in Mathematical Physics}\ }\textbf {\bibinfo
  {volume} {144}},\ \bibinfo {pages} {443} (\bibinfo {year}
  {1992})}\BibitemShut {NoStop}%
\bibitem [{\citenamefont {{Klumper}}\ \emph {et~al.}(1991)\citenamefont
  {{Klumper}}, \citenamefont {{Schadschneider}},\ and\ \citenamefont
  {{Zittartz}}}]{MPS_2}%
  \BibitemOpen
  \bibfield  {author} {\bibinfo {author} {\bibfnamefont {A.}~\bibnamefont
  {{Klumper}}}, \bibinfo {author} {\bibfnamefont {A.}~\bibnamefont
  {{Schadschneider}}}, \ and\ \bibinfo {author} {\bibfnamefont
  {J.}~\bibnamefont {{Zittartz}}},\ }\href {\doibase
  https://doi.org/10.1088/0305-4470/24/16/012} {\bibfield  {journal} {\bibinfo
  {journal} {Journal of Physics A Mathematical General}\ }\textbf {\bibinfo
  {volume} {24}},\ \bibinfo {pages} {L955} (\bibinfo {year}
  {1991})}\BibitemShut {NoStop}%
\bibitem [{\citenamefont {{Kl{\"u}mper}}\ \emph {et~al.}(1993)\citenamefont
  {{Kl{\"u}mper}}, \citenamefont {{Schadschneider}},\ and\ \citenamefont
  {{Zittartz}}}]{MPS_3}%
  \BibitemOpen
  \bibfield  {author} {\bibinfo {author} {\bibfnamefont {A.}~\bibnamefont
  {{Kl{\"u}mper}}}, \bibinfo {author} {\bibfnamefont {A.}~\bibnamefont
  {{Schadschneider}}}, \ and\ \bibinfo {author} {\bibfnamefont
  {J.}~\bibnamefont {{Zittartz}}},\ }\href {\doibase
  https://doi.org/10.1209/0295-5075/24/4/010} {\bibfield  {journal} {\bibinfo
  {journal} {EPL (Europhysics Letters)}\ }\textbf {\bibinfo {volume} {24}},\
  \bibinfo {pages} {293} (\bibinfo {year} {1993})}\BibitemShut {NoStop}%
\bibitem [{\citenamefont {Verstraete}\ \emph {et~al.}(2006)\citenamefont
  {Verstraete}, \citenamefont {Wolf}, \citenamefont {Perez-Garcia},\ and\
  \citenamefont {Cirac}}]{PhysRevLett.96.220601}%
  \BibitemOpen
  \bibfield  {author} {\bibinfo {author} {\bibfnamefont {F.}~\bibnamefont
  {Verstraete}}, \bibinfo {author} {\bibfnamefont {M.~M.}\ \bibnamefont
  {Wolf}}, \bibinfo {author} {\bibfnamefont {D.}~\bibnamefont {Perez-Garcia}},
  \ and\ \bibinfo {author} {\bibfnamefont {J.~I.}\ \bibnamefont {Cirac}},\
  }\href {\doibase https://doi.org/10.1103/PhysRevLett.96.220601} {\bibfield
  {journal} {\bibinfo  {journal} {Phys. Rev. Lett.}\ }\textbf {\bibinfo
  {volume} {96}},\ \bibinfo {pages} {220601} (\bibinfo {year}
  {2006})}\BibitemShut {NoStop}%
\bibitem [{\citenamefont {Verstraete}\ and\ \citenamefont
  {Cirac}(2004)}]{2004cond.mat.7066V}%
  \BibitemOpen
  \bibfield  {author} {\bibinfo {author} {\bibfnamefont {F.}~\bibnamefont
  {Verstraete}}\ and\ \bibinfo {author} {\bibfnamefont {J.~I.}\ \bibnamefont
  {Cirac}},\ }\href {https://arxiv.org/abs/cond-mat/0407066} {\bibfield
  {journal} {\bibinfo  {journal} {arXiv preprint cond-mat/0407066}\ } (\bibinfo
  {year} {2004})}\BibitemShut {NoStop}%
\bibitem [{\citenamefont {Tepaske}\ and\ \citenamefont
  {Luitz}(2021)}]{tepaske2020threedimensional}%
  \BibitemOpen
  \bibfield  {author} {\bibinfo {author} {\bibfnamefont {M.~S.~J.}\
  \bibnamefont {Tepaske}}\ and\ \bibinfo {author} {\bibfnamefont {D.~J.}\
  \bibnamefont {Luitz}},\ }\href {\doibase
  https://doi.org/10.1103/PhysRevResearch.3.023236} {\bibfield  {journal}
  {\bibinfo  {journal} {Phys. Rev. Research}\ }\textbf {\bibinfo {volume}
  {3}},\ \bibinfo {pages} {023236} (\bibinfo {year} {2021})}\BibitemShut
  {NoStop}%
\bibitem [{\citenamefont {Vidal}(2007)}]{PhysRevLett.99.220405}%
  \BibitemOpen
  \bibfield  {author} {\bibinfo {author} {\bibfnamefont {G.}~\bibnamefont
  {Vidal}},\ }\href {\doibase https://doi.org/10.1103/PhysRevLett.99.220405}
  {\bibfield  {journal} {\bibinfo  {journal} {Phys. Rev. Lett.}\ }\textbf
  {\bibinfo {volume} {99}},\ \bibinfo {pages} {220405} (\bibinfo {year}
  {2007})}\BibitemShut {NoStop}%
\bibitem [{\citenamefont {Evenbly}\ and\ \citenamefont
  {Vidal}(2009)}]{PhysRevLett.102.180406}%
  \BibitemOpen
  \bibfield  {author} {\bibinfo {author} {\bibfnamefont {G.}~\bibnamefont
  {Evenbly}}\ and\ \bibinfo {author} {\bibfnamefont {G.}~\bibnamefont
  {Vidal}},\ }\href {\doibase https://doi.org/10.1103/PhysRevLett.102.180406}
  {\bibfield  {journal} {\bibinfo  {journal} {Phys. Rev. Lett.}\ }\textbf
  {\bibinfo {volume} {102}},\ \bibinfo {pages} {180406} (\bibinfo {year}
  {2009})}\BibitemShut {NoStop}%
\bibitem [{\citenamefont {Shi}\ \emph {et~al.}(2006)\citenamefont {Shi},
  \citenamefont {Duan},\ and\ \citenamefont {Vidal}}]{PhysRevA.74.022320}%
  \BibitemOpen
  \bibfield  {author} {\bibinfo {author} {\bibfnamefont {Y.-Y.}\ \bibnamefont
  {Shi}}, \bibinfo {author} {\bibfnamefont {L.-M.}\ \bibnamefont {Duan}}, \
  and\ \bibinfo {author} {\bibfnamefont {G.}~\bibnamefont {Vidal}},\ }\href
  {\doibase https://doi.org/10.1103/PhysRevA.74.022320} {\bibfield  {journal}
  {\bibinfo  {journal} {Phys. Rev. A}\ }\textbf {\bibinfo {volume} {74}},\
  \bibinfo {pages} {022320} (\bibinfo {year} {2006})}\BibitemShut {NoStop}%
\bibitem [{\citenamefont {Gerster}\ \emph {et~al.}(2014)\citenamefont
  {Gerster}, \citenamefont {Silvi}, \citenamefont {Rizzi}, \citenamefont
  {Fazio}, \citenamefont {Calarco},\ and\ \citenamefont
  {Montangero}}]{PhysRevB.90.125154}%
  \BibitemOpen
  \bibfield  {author} {\bibinfo {author} {\bibfnamefont {M.}~\bibnamefont
  {Gerster}}, \bibinfo {author} {\bibfnamefont {P.}~\bibnamefont {Silvi}},
  \bibinfo {author} {\bibfnamefont {M.}~\bibnamefont {Rizzi}}, \bibinfo
  {author} {\bibfnamefont {R.}~\bibnamefont {Fazio}}, \bibinfo {author}
  {\bibfnamefont {T.}~\bibnamefont {Calarco}}, \ and\ \bibinfo {author}
  {\bibfnamefont {S.}~\bibnamefont {Montangero}},\ }\href {\doibase
  https://doi.org/10.1103/PhysRevB.90.125154} {\bibfield  {journal} {\bibinfo
  {journal} {Phys. Rev. B}\ }\textbf {\bibinfo {volume} {90}},\ \bibinfo
  {pages} {125154} (\bibinfo {year} {2014})}\BibitemShut {NoStop}%
\bibitem [{\citenamefont {Silvi}\ \emph {et~al.}(2010)\citenamefont {Silvi},
  \citenamefont {Giovannetti}, \citenamefont {Montangero}, \citenamefont
  {Rizzi}, \citenamefont {Cirac},\ and\ \citenamefont
  {Fazio}}]{PhysRevA.81.062335}%
  \BibitemOpen
  \bibfield  {author} {\bibinfo {author} {\bibfnamefont {P.}~\bibnamefont
  {Silvi}}, \bibinfo {author} {\bibfnamefont {V.}~\bibnamefont {Giovannetti}},
  \bibinfo {author} {\bibfnamefont {S.}~\bibnamefont {Montangero}}, \bibinfo
  {author} {\bibfnamefont {M.}~\bibnamefont {Rizzi}}, \bibinfo {author}
  {\bibfnamefont {J.~I.}\ \bibnamefont {Cirac}}, \ and\ \bibinfo {author}
  {\bibfnamefont {R.}~\bibnamefont {Fazio}},\ }\href {\doibase
  https://doi.org/10.1103/PhysRevA.81.062335} {\bibfield  {journal} {\bibinfo
  {journal} {Phys. Rev. A}\ }\textbf {\bibinfo {volume} {81}},\ \bibinfo
  {pages} {062335} (\bibinfo {year} {2010})}\BibitemShut {NoStop}%
\bibitem [{\citenamefont {Silvi}\ \emph {et~al.}(2019)\citenamefont {Silvi},
  \citenamefont {Tschirsich}, \citenamefont {Gerster}, \citenamefont
  {Junemann}, \citenamefont {Jaschke}, \citenamefont {Rizzi},\ and\
  \citenamefont {Montangero}}]{TN_Anthology}%
  \BibitemOpen
  \bibfield  {author} {\bibinfo {author} {\bibfnamefont {P.}~\bibnamefont
  {Silvi}}, \bibinfo {author} {\bibfnamefont {F.}~\bibnamefont {Tschirsich}},
  \bibinfo {author} {\bibfnamefont {M.}~\bibnamefont {Gerster}}, \bibinfo
  {author} {\bibfnamefont {J.}~\bibnamefont {Junemann}}, \bibinfo {author}
  {\bibfnamefont {D.}~\bibnamefont {Jaschke}}, \bibinfo {author} {\bibfnamefont
  {M.}~\bibnamefont {Rizzi}}, \ and\ \bibinfo {author} {\bibfnamefont
  {S.}~\bibnamefont {Montangero}},\ }\href {\doibase
  https://doi.org/10.21468/SciPostPhysLectNotes.8} {\bibfield  {journal}
  {\bibinfo  {journal} {SciPost Phys. Lect. Notes}\ ,\ \bibinfo {pages} {8}}
  (\bibinfo {year} {2019})}\BibitemShut {NoStop}%
\bibitem [{\citenamefont {Felser}\ \emph {et~al.}(2021)\citenamefont {Felser},
  \citenamefont {Notarnicola},\ and\ \citenamefont {Montangero}}]{aTTN_2021}%
  \BibitemOpen
  \bibfield  {author} {\bibinfo {author} {\bibfnamefont {T.}~\bibnamefont
  {Felser}}, \bibinfo {author} {\bibfnamefont {S.}~\bibnamefont {Notarnicola}},
  \ and\ \bibinfo {author} {\bibfnamefont {S.}~\bibnamefont {Montangero}},\
  }\href {\doibase 10.1103/PhysRevLett.126.170603} {\bibfield  {journal}
  {\bibinfo  {journal} {Phys. Rev. Lett.}\ }\textbf {\bibinfo {volume} {126}},\
  \bibinfo {pages} {170603} (\bibinfo {year} {2021})}\BibitemShut {NoStop}%
\bibitem [{\citenamefont {Qian}\ and\ \citenamefont
  {Qin}(2021)}]{qian2021tree}%
  \BibitemOpen
  \bibfield  {author} {\bibinfo {author} {\bibfnamefont {X.}~\bibnamefont
  {Qian}}\ and\ \bibinfo {author} {\bibfnamefont {M.}~\bibnamefont {Qin}},\
  }\href@noop {} {\  (\bibinfo {year} {2021})},\ \Eprint
  {http://arxiv.org/abs/2110.08794} {arXiv:2110.08794 [quant-ph]} \BibitemShut
  {NoStop}%
\bibitem [{\citenamefont {Magnifico}\ \emph {et~al.}(2021)\citenamefont
  {Magnifico}, \citenamefont {Felser}, \citenamefont {Silvi},\ and\
  \citenamefont {Montangero}}]{magnifico2020lattice}%
  \BibitemOpen
  \bibfield  {author} {\bibinfo {author} {\bibfnamefont {G.}~\bibnamefont
  {Magnifico}}, \bibinfo {author} {\bibfnamefont {T.}~\bibnamefont {Felser}},
  \bibinfo {author} {\bibfnamefont {P.}~\bibnamefont {Silvi}}, \ and\ \bibinfo
  {author} {\bibfnamefont {S.}~\bibnamefont {Montangero}},\ }\href {\doibase
  10.1038/s41467-021-23646-3} {\bibfield  {journal} {\bibinfo  {journal}
  {Nature Communications}\ }\textbf {\bibinfo {volume} {12}},\ \bibinfo {pages}
  {3600} (\bibinfo {year} {2021})}\BibitemShut {NoStop}%
\bibitem [{\citenamefont {Hikihara}\ \emph {et~al.}(1999)\citenamefont
  {Hikihara}, \citenamefont {Furusaki},\ and\ \citenamefont
  {Sigrist}}]{PhysRevB.60.12116}%
  \BibitemOpen
  \bibfield  {author} {\bibinfo {author} {\bibfnamefont {T.}~\bibnamefont
  {Hikihara}}, \bibinfo {author} {\bibfnamefont {A.}~\bibnamefont {Furusaki}},
  \ and\ \bibinfo {author} {\bibfnamefont {M.}~\bibnamefont {Sigrist}},\ }\href
  {\doibase 10.1103/PhysRevB.60.12116} {\bibfield  {journal} {\bibinfo
  {journal} {Phys. Rev. B}\ }\textbf {\bibinfo {volume} {60}},\ \bibinfo
  {pages} {12116} (\bibinfo {year} {1999})}\BibitemShut {NoStop}%
\bibitem [{\citenamefont {Goldsborough}\ and\ \citenamefont
  {R\"omer}(2014)}]{PhysRevB.89.214203}%
  \BibitemOpen
  \bibfield  {author} {\bibinfo {author} {\bibfnamefont {A.~M.}\ \bibnamefont
  {Goldsborough}}\ and\ \bibinfo {author} {\bibfnamefont {R.~A.}\ \bibnamefont
  {R\"omer}},\ }\href {\doibase 10.1103/PhysRevB.89.214203} {\bibfield
  {journal} {\bibinfo  {journal} {Phys. Rev. B}\ }\textbf {\bibinfo {volume}
  {89}},\ \bibinfo {pages} {214203} (\bibinfo {year} {2014})}\BibitemShut
  {NoStop}%
\bibitem [{\citenamefont {Lin}\ \emph {et~al.}(2017)\citenamefont {Lin},
  \citenamefont {Kao}, \citenamefont {Chen},\ and\ \citenamefont
  {Lin}}]{PhysRevB.96.064427}%
  \BibitemOpen
  \bibfield  {author} {\bibinfo {author} {\bibfnamefont {Y.-P.}\ \bibnamefont
  {Lin}}, \bibinfo {author} {\bibfnamefont {Y.-J.}\ \bibnamefont {Kao}},
  \bibinfo {author} {\bibfnamefont {P.}~\bibnamefont {Chen}}, \ and\ \bibinfo
  {author} {\bibfnamefont {Y.-C.}\ \bibnamefont {Lin}},\ }\href {\doibase
  10.1103/PhysRevB.96.064427} {\bibfield  {journal} {\bibinfo  {journal} {Phys.
  Rev. B}\ }\textbf {\bibinfo {volume} {96}},\ \bibinfo {pages} {064427}
  (\bibinfo {year} {2017})}\BibitemShut {NoStop}%
\bibitem [{\citenamefont {Tsai}\ \emph {et~al.}(2020)\citenamefont {Tsai},
  \citenamefont {Chen},\ and\ \citenamefont {Lin}}]{Tsai_2020}%
  \BibitemOpen
  \bibfield  {author} {\bibinfo {author} {\bibfnamefont {Z.-L.}\ \bibnamefont
  {Tsai}}, \bibinfo {author} {\bibfnamefont {P.}~\bibnamefont {Chen}}, \ and\
  \bibinfo {author} {\bibfnamefont {Y.-C.}\ \bibnamefont {Lin}},\ }\href
  {https://doi.org/10.1140/epjb/e2020-100585-8} {\bibfield  {journal} {\bibinfo
   {journal} {The European Physical Journal B}\ }\textbf {\bibinfo {volume}
  {93}} (\bibinfo {year} {2020})}\BibitemShut {NoStop}%
\bibitem [{\citenamefont {Seki}\ \emph {et~al.}(2020)\citenamefont {Seki},
  \citenamefont {Hikihara},\ and\ \citenamefont {Okunishi}}]{Seki_2020}%
  \BibitemOpen
  \bibfield  {author} {\bibinfo {author} {\bibfnamefont {K.}~\bibnamefont
  {Seki}}, \bibinfo {author} {\bibfnamefont {T.}~\bibnamefont {Hikihara}}, \
  and\ \bibinfo {author} {\bibfnamefont {K.}~\bibnamefont {Okunishi}},\ }\href
  {\doibase 10.1103/PhysRevB.102.144439} {\bibfield  {journal} {\bibinfo
  {journal} {Phys. Rev. B}\ }\textbf {\bibinfo {volume} {102}},\ \bibinfo
  {pages} {144439} (\bibinfo {year} {2020})}\BibitemShut {NoStop}%
\bibitem [{\citenamefont {Seki}\ \emph {et~al.}(2021)\citenamefont {Seki},
  \citenamefont {Hikihara},\ and\ \citenamefont
  {Okunishi}}]{seki2021entanglementbased}%
  \BibitemOpen
  \bibfield  {author} {\bibinfo {author} {\bibfnamefont {K.}~\bibnamefont
  {Seki}}, \bibinfo {author} {\bibfnamefont {T.}~\bibnamefont {Hikihara}}, \
  and\ \bibinfo {author} {\bibfnamefont {K.}~\bibnamefont {Okunishi}},\ }\href
  {\doibase 10.1103/PhysRevB.104.134405} {\bibfield  {journal} {\bibinfo
  {journal} {Phys. Rev. B}\ }\textbf {\bibinfo {volume} {104}},\ \bibinfo
  {pages} {134405} (\bibinfo {year} {2021})}\BibitemShut {NoStop}%
\bibitem [{\citenamefont {Cataldi}\ \emph {et~al.}(2021)\citenamefont
  {Cataldi}, \citenamefont {Abedi}, \citenamefont {Magnifico}, \citenamefont
  {Notarnicola}, \citenamefont {Pozza}, \citenamefont {Giovannetti},\ and\
  \citenamefont {Montangero}}]{Cataldi_2021}%
  \BibitemOpen
  \bibfield  {author} {\bibinfo {author} {\bibfnamefont {G.}~\bibnamefont
  {Cataldi}}, \bibinfo {author} {\bibfnamefont {A.}~\bibnamefont {Abedi}},
  \bibinfo {author} {\bibfnamefont {G.}~\bibnamefont {Magnifico}}, \bibinfo
  {author} {\bibfnamefont {S.}~\bibnamefont {Notarnicola}}, \bibinfo {author}
  {\bibfnamefont {N.~D.}\ \bibnamefont {Pozza}}, \bibinfo {author}
  {\bibfnamefont {V.}~\bibnamefont {Giovannetti}}, \ and\ \bibinfo {author}
  {\bibfnamefont {S.}~\bibnamefont {Montangero}},\ }\href {\doibase
  10.22331/q-2021-09-29-556} {\bibfield  {journal} {\bibinfo  {journal}
  {Quantum}\ }\textbf {\bibinfo {volume} {5}},\ \bibinfo {pages} {556}
  (\bibinfo {year} {2021})}\BibitemShut {NoStop}%
\bibitem [{\citenamefont {Dutta}\ \emph {et~al.}(2015)\citenamefont {Dutta},
  \citenamefont {Aeppli}, \citenamefont {Chakrabarti}, \citenamefont
  {Divakaran}, \citenamefont {Rosenbaum},\ and\ \citenamefont
  {Sen}}]{dutta2015quantum}%
  \BibitemOpen
  \bibfield  {author} {\bibinfo {author} {\bibfnamefont {A.}~\bibnamefont
  {Dutta}}, \bibinfo {author} {\bibfnamefont {G.}~\bibnamefont {Aeppli}},
  \bibinfo {author} {\bibfnamefont {B.~K.}\ \bibnamefont {Chakrabarti}},
  \bibinfo {author} {\bibfnamefont {U.}~\bibnamefont {Divakaran}}, \bibinfo
  {author} {\bibfnamefont {T.~F.}\ \bibnamefont {Rosenbaum}}, \ and\ \bibinfo
  {author} {\bibfnamefont {D.}~\bibnamefont {Sen}},\ }\href@noop {} {\
  (\bibinfo {year} {2015})},\ \Eprint {http://arxiv.org/abs/1012.0653}
  {arXiv:1012.0653 [cond-mat.stat-mech]} \BibitemShut {NoStop}%
\bibitem [{\citenamefont {Rieger}\ and\ \citenamefont
  {Young}(1994)}]{PhysRevLett.72.4141}%
  \BibitemOpen
  \bibfield  {author} {\bibinfo {author} {\bibfnamefont {H.}~\bibnamefont
  {Rieger}}\ and\ \bibinfo {author} {\bibfnamefont {A.~P.}\ \bibnamefont
  {Young}},\ }\href {\doibase 10.1103/PhysRevLett.72.4141} {\bibfield
  {journal} {\bibinfo  {journal} {Phys. Rev. Lett.}\ }\textbf {\bibinfo
  {volume} {72}},\ \bibinfo {pages} {4141} (\bibinfo {year}
  {1994})}\BibitemShut {NoStop}%
\bibitem [{\citenamefont {Laflorencie}(2016)}]{area_law_2016}%
  \BibitemOpen
  \bibfield  {author} {\bibinfo {author} {\bibfnamefont {N.}~\bibnamefont
  {Laflorencie}},\ }\href {\doibase 10.1016/j.physrep.2016.06.008} {\bibfield
  {journal} {\bibinfo  {journal} {Physics Reports}\ }\textbf {\bibinfo {volume}
  {646}},\ \bibinfo {pages} {1} (\bibinfo {year} {2016})}\BibitemShut {NoStop}%
\bibitem [{\citenamefont {Nielsen}\ and\ \citenamefont
  {Chuang}(2010)}]{nielsen_chuang_2010}%
  \BibitemOpen
  \bibfield  {author} {\bibinfo {author} {\bibfnamefont {M.~A.}\ \bibnamefont
  {Nielsen}}\ and\ \bibinfo {author} {\bibfnamefont {I.~L.}\ \bibnamefont
  {Chuang}},\ }\href {\doibase 10.1017/CBO9780511976667} {\emph {\bibinfo
  {title} {Quantum Computation and Quantum Information: 10th Anniversary
  Edition}}}\ (\bibinfo  {publisher} {Cambridge University Press},\ \bibinfo
  {year} {2010})\BibitemShut {NoStop}%
\bibitem [{\citenamefont {Alba}\ \emph {et~al.}(2019)\citenamefont {Alba},
  \citenamefont {Santalla}, \citenamefont {Ruggiero}, \citenamefont
  {Rodriguez-Laguna}, \citenamefont {Calabrese},\ and\ \citenamefont
  {Sierra}}]{area_law_violation_2019}%
  \BibitemOpen
  \bibfield  {author} {\bibinfo {author} {\bibfnamefont {V.}~\bibnamefont
  {Alba}}, \bibinfo {author} {\bibfnamefont {S.~N.}\ \bibnamefont {Santalla}},
  \bibinfo {author} {\bibfnamefont {P.}~\bibnamefont {Ruggiero}}, \bibinfo
  {author} {\bibfnamefont {J.}~\bibnamefont {Rodriguez-Laguna}}, \bibinfo
  {author} {\bibfnamefont {P.}~\bibnamefont {Calabrese}}, \ and\ \bibinfo
  {author} {\bibfnamefont {G.}~\bibnamefont {Sierra}},\ }\href {\doibase
  10.1088/1742-5468/ab02df} {\bibfield  {journal} {\bibinfo  {journal} {Journal
  of Statistical Mechanics: Theory and Experiment}\ }\textbf {\bibinfo {volume}
  {2019}},\ \bibinfo {pages} {023105} (\bibinfo {year} {2019})}\BibitemShut
  {NoStop}%
\bibitem [{\citenamefont {Yu}\ \emph {et~al.}(2008)\citenamefont {Yu},
  \citenamefont {Saleur},\ and\ \citenamefont
  {Haas}}]{disorder_ising_area_law_2008}%
  \BibitemOpen
  \bibfield  {author} {\bibinfo {author} {\bibfnamefont {R.}~\bibnamefont
  {Yu}}, \bibinfo {author} {\bibfnamefont {H.}~\bibnamefont {Saleur}}, \ and\
  \bibinfo {author} {\bibfnamefont {S.}~\bibnamefont {Haas}},\ }\href {\doibase
  10.1103/PhysRevB.77.140402} {\bibfield  {journal} {\bibinfo  {journal} {Phys.
  Rev. B}\ }\textbf {\bibinfo {volume} {77}},\ \bibinfo {pages} {140402}
  (\bibinfo {year} {2008})}\BibitemShut {NoStop}%
\bibitem [{\citenamefont {Kov\'acs}\ and\ \citenamefont
  {Igl\'oi}(2012)}]{disorder_ising_area_law_2012}%
  \BibitemOpen
  \bibfield  {author} {\bibinfo {author} {\bibfnamefont {I.~A.}\ \bibnamefont
  {Kov\'acs}}\ and\ \bibinfo {author} {\bibfnamefont {F.}~\bibnamefont
  {Igl\'oi}},\ }\href {\doibase 10.1209/0295-5075/97/67009} {\bibfield
  {journal} {\bibinfo  {journal} {EPL (Europhysics Letters)}\ }\textbf
  {\bibinfo {volume} {97}},\ \bibinfo {pages} {67009} (\bibinfo {year}
  {2012})}\BibitemShut {NoStop}%
\bibitem [{\citenamefont {Nakatani}\ and\ \citenamefont
  {Chan}(2013)}]{doi:10.1063/1.4798639}%
  \BibitemOpen
  \bibfield  {author} {\bibinfo {author} {\bibfnamefont {N.}~\bibnamefont
  {Nakatani}}\ and\ \bibinfo {author} {\bibfnamefont {G.~K.-L.}\ \bibnamefont
  {Chan}},\ }\href {\doibase 10.1063/1.4798639} {\bibfield  {journal} {\bibinfo
   {journal} {The Journal of Chemical Physics}\ }\textbf {\bibinfo {volume}
  {138}},\ \bibinfo {pages} {134113} (\bibinfo {year} {2013})}\BibitemShut
  {NoStop}%
\bibitem [{\citenamefont {Lehoucq}\ \emph {et~al.}(1997)\citenamefont
  {Lehoucq}, \citenamefont {Sorensen},\ and\ \citenamefont
  {Yang}}]{Lehoucq97arpackusers}%
  \BibitemOpen
  \bibfield  {author} {\bibinfo {author} {\bibfnamefont {R.~B.}\ \bibnamefont
  {Lehoucq}}, \bibinfo {author} {\bibfnamefont {D.~C.}\ \bibnamefont
  {Sorensen}}, \ and\ \bibinfo {author} {\bibfnamefont {C.}~\bibnamefont
  {Yang}},\ }\href@noop {} {\enquote {\bibinfo {title} {Arpack users guide:
  Solution of large scale eigenvalue problems by implicitly restarted arnoldi
  methods.}}\ } (\bibinfo {year} {1997})\BibitemShut {NoStop}%
\end{thebibliography}

%

\end{document}